\documentclass[twocolumn]{aastex61}
\usepackage{silence}
\usepackage{graphicx}
\graphicspath{{images/}}
\usepackage[caption=false]{subfig}
\usepackage{booktabs}

\usepackage{natbib}
\setcitestyle{authoryear,open={(},close={)}}

\usepackage{float}
\usepackage{color,soul}
\usepackage{wrapfig}
\usepackage{bm}
\usepackage{amsmath}

\usepackage{fancyhdr}
\usepackage{wasysym}

\newcommand{\cii}{[\ion{C}{2}]}
\newcommand{\oi}{[\ion{O}{1}]}
\newcommand{\Htwo}{H$_2$}
\newcommand{\mathcii}{[\mbox{\footnotesize \ion{C}{2}}]}
\newcommand{\lir}{L$_{\mathrm{IR}}$}
\newcommand{\lcii}{L$_{[\mbox{\footnotesize \ion{C}{2}}]}$}
\newcommand{\loi}{L$_{[\mbox{\footnotesize \ion{O}{1}}]}$}
\newcommand{\mathlcii}{ \mathrm{L}_{[\mbox{\footnotesize \ion{C}{2}}]}}
\newcommand{\loglir}{$\log\mathrm{L}_{\mathrm{IR}}/\mathrm{L}_\odot$}
\newcommand{\sigmaIR}{$\Sigma_{\mathrm{IR}}$}
\newcommand{\sigmaSFR}{$\Sigma_{\mathrm{SFR}}$}
\newcommand{\mum}{$\mu$m}

\newcommand{\loglc}{$\log\mathrm{L}_{[\mathrm{\tiny C\ II]}}$}
\newcommand{\lpah}{L$_{\mathrm{\tiny PAH}}$}
\newcommand{\lsix}{L$_{6.2\mu\mathrm{\tiny m}}$}
\newcommand{\lii}{L$_{11.3\mu\mathrm{\tiny m}}$}

\newcommand{\kms}{km s$^{-1}$}
\newcommand{\zirs}{$z_{\mathrm{\tiny PAH}}$}
\newcommand{\ewPAH}{EW$_{6.2}$}
\newcommand{\hband}{\textit{H}$_{160}$}

\newcommand{\ephot}{$\epsilon_{\tiny\mathrm{PE}}$}
\newcommand{\fagn}{$f_{\tiny\mathrm{AGN,MIR}}$}
\newcommand{\lcold}{L$_{\tiny\mathrm{cold}}$}
\newcommand{\zcii}{$z_{[\mbox{\footnotesize \ion{C}{2}}]}$}
\newcommand{\Reff}{$R_{\mbox{\footnotesize eff},160}$}

\begin{document}

\title{Measuring the heating and cooling of the interstellar medium at high redshift: PAH and \cii\ observations of the same star forming galaxies at $z\sim2$}
\author{Jed McKinney} 
\affiliation{Department of Astronomy, University of Massachusetts, Amherst, MA 01003, USA.}
\email{jhmckinney@umass.edu}
\author{Alexandra Pope}
\affiliation{Department of Astronomy, University of Massachusetts, Amherst, MA 01003, USA.}
\author{Lee Armus}
\affiliation{Infrared Processing and Analysis Center, MC 314-6, Caltech, 1200 E. California Blvd., Pasadena, CA 91125, USA.}
\author{Ranga-Ram Chary}
\affiliation{Infrared Processing and Analysis Center, MC 314-6, Caltech, 1200 E. California Blvd., Pasadena, CA 91125, USA.}
\author{Tanio D\'iaz-Santos}
\affiliation{N\'ucleo de Astronom\'ia de la Facultad de Ingenier\'ia y Ciencias, Universidad Diego Portales, Av. Ej\'ercito Libertador 441, Santiago, Chile}
\affiliation{Chinese Academy of Sciences South America Center for Astronomy, National Astronomical Observatories, CAS, Beijing 100101, China}
\author{Mark E. Dickinson}
\affiliation{National Optical Astronomy Observatory, 950 North Cherry Avenue, Tucson, AZ 85719, USA.}
\author{Allison Kirkpatrick}
\affiliation{Department of Physics \& Astronomy, University of Kansas, Lawrence, KS 66045, USA}

\begin{abstract}
    Star formation depends critically on cooling mechanisms in the interstellar medium (ISM); however, thermal properties of gas in galaxies at the peak epoch of star formation ($z\sim2$) remain poorly understood. A limiting factor in understanding the multiphase ISM is the lack of multiple tracers detected in the same galaxies, such as Polycyclic Aromatic Hydrocarbon (PAH) emission, a tracer of a critical photoelectric heating mechanism in interstellar gas, and \cii\ 158\mum\ fine-structure emission, a principal coolant. We present ALMA Band 9 observations targeting \cii\ in six $z\sim2$ star-forming galaxies with strong \textit{Spitzer} IRS detections of PAH emission. All six galaxies are detected in dust continuum and marginally resolved. We compare the properties of PAH and \cii\ emission, and constrain their relationship as a function of total infrared luminosity (\lir) and IR surface density. \cii\ emission is detected in one galaxy at high signal-to-noise (34$\sigma$), and we place a secure upper limit on a second source. The rest of our sample are not detected in \cii\ likely due to redshift uncertainties and narrow ALMA bandpass windows. Our results are consistent with the deficit in \cii/\lir\ and PAH/\lir\ observed in the literature. However, the ratio of \cii\ to PAH emission at $z\sim2$ is possibly much lower than what is observed in nearby dusty star-forming galaxies. This could be the result of enhanced cooling via \oi\ at high$-z$, hotter gas and dust temperatures, and/or a reduction in the photoelectric efficiency, in which the coupling between interstellar radiation and gas heating is diminished. 
\end{abstract}

\section{Introduction}
Ten billion years ago ($z\sim2$), the star-formation rate density of the Universe peaked and individual galaxies were forming more stars than at any other time in cosmic history (e.g., \citealt{Lilly1996,Madau1996,Chary2001,MadauDickinson2014}). Enhanced star-formation was promoted by gas resupply through cold mode accretion onto galaxies (e.g., \citealt{Keres2005,Keres2009,Genzel2008,Tacconi2010}), accompanied by a change in the efficiency of star-formation (e.g., \citealt{Tacconi2010,Tacconi2013,Tacconi2018,Genzel2015,Scoville2017,Liu2019}), which suggests evolution in the heating and cooling mechanisms of interstellar gas. 

The internal transfer of thermal energy is critical for any physical system. Photoelectrons ejected from polycyclic aromatic hydrocarbons (PAHs) are thought to be the most important, albeit inefficient, mechanism for converting stellar radiation to thermal energy in and around sites of active star-formation \citep{Watson1972,Bakes1994,Helou2001}. PAH molecules are complex grains comprised mostly of C and H, and are common in photodissociation regions (PDRs) where gas densities of $n\sim10^3-10^6$ cm$^{-3}$ are illuminated by far-UV stellar radiation fields \citep{TH1985}. Once excited by stellar photons, PAHs emit vibrational lines between $5-15\ \mu$m that can contain as much as $\sim20\%$ of total IR emission (\lir, $8-1000$ $\mu$m) \citep{Smith2007,Sajina2007,Pope2008,Dale2009}. Therefore, mid-IR PAH features are direct probes of photoelectric heating in dense PDRs and a key diagnostic of the interstellar medium (ISM). 

The energy injected into the ISM by photoelectrons is radiated away in the infrared (IR). Far-IR fine-structure emission lines such as \cii\ at 158 $\mu$m and \oi\ at 63 $\mu$m can contain $0.1-1\%$ of \lir\ \citep{TH1985,Stacey2010,DiazSantos2013,Brisbin2015,Ibar2015}. \cii\ in particular is emerging as a powerful, but complicated diagnostic of the ISM because it comes from different regions in a galaxy. With a critical density of $n_{\mathrm{\tiny crit}}\sim3\times10^3-6\times10^3$ cm$^{-3}$ at $\sim100$ K, \cii\ is collisionally excited by H and \Htwo\ in PDRs, as well as by warm electrons at 8,000 K \citep{Goldsmith2012}. Ancillary observations of [\ion{N}{2}] 205 $\mu$m emission constrain the fraction of \cii\ emission originating from PDRs (e.g., \citealt{Croxall2012}), {which is greater for lower metallcities \citep{Croxall2017,Cormier2019}, and approaches unity in warm and compact, dusty, star forming regions \citep{Sutter2019}. Thus, \cii\ can be used to trace PDR cooling in warm, compact environments, a critical physical process in atomic gas for  star-formation to occur. 


\begin{deluxetable*}{lccccccccc}
    \tablecaption{Sample Summary\label{sampsum}}
    \tabletypesize{\footnotesize}
    \tablehead{Target & R.A. & Dec.  & $z_{\mathrm{\tiny IRS}}$ &  $\log\,\mathrm{L}_{\mathrm{IR}}$ & $\log\,$\lsix & $\log\,$\lii & $\log\,$M$_*$\tablenotemark{a} & SFR$_{\tiny\mathrm{IR}}$ & \fagn\tablenotemark{b} \\ & [J2000] & [J2000] & & [L$_\odot$] & [L$_\odot$] & [L$_\odot$] & [M$_\odot$] & [M$_\odot$ yr$^{-1}$]
    }
    \startdata 
    GS IRS20 & 03:32:47.58 & -27:44:52.0 & $1.923^{+0.030}_{-0.030}$ & $13.06\pm0.12$ & $9.99^{+0.12}_{-0.12}$  & $10.11^{+0.10}_{-0.10}$ & 10.98 & 717 & $0.2$ \\[.3ex]
    GS IRS46 & 03:32:42.71 & -27:39:27.0 & $1.850^{+0.014}_{-0.011}$ & $12.63\pm0.29$ & $9.90^{+0.15}_{-0.15}$  & - \tablenotemark{c} & - \tablenotemark{d} & 376 & $0.0$ \\[.3ex]
    GS IRS50 & 03:32:31.52 & -27:48:53.0 & $1.900^{+0.081}_{-0.041}$ & $12.46\pm0.15$ & $10.17^{+0.09}_{-0.09}$ & $9.66^{+0.33}_{-0.33}$  & 11.03 & 184 & $0.0$ \\[.3ex]
    GS IRS52 & 03:32:12.52 & -27:43:06.0 & $1.824^{+0.018}_{-0.020}$ & $12.53\pm0.29$ & $9.91^{+0.12}_{-0.12}$  & $10.10^{+0.13}_{-0.13}$ & 10.64 & 232 & $0.0$ \\[.3ex]
    GS IRS58 & 03:32:40.24 & -27:49:49.0 & $1.890^{+0.017}_{-0.042}$ & $12.52\pm0.17$ & $9.91^{+0.10}_{-0.10}$  & $9.96^{+0.25}_{-0.25}$  & 11.07 & 207 & $0.0$\\[.3ex]
    GS IRS61 & 03:32:43.45 & -27:49:01.0 & $1.759^{+0.016}_{-0.008}$ & $12.46\pm0.13$ & $10.18^{+0.04}_{-0.04}$ & $9.75^{+0.10}_{-0.10}$  & 10.90 & 243 & $0.0$ \\[.3ex]
    \enddata
    \tablecomments{When calculating M$_*$ and SFR$_{\tiny\mathrm{IR}}$, we assume a Salpeter IMF and SFR$_{\tiny\mathrm{IR}} \approx 1.8\times10^{-10}\,$\lir\, \citep{Kennicutt1998}. We assume a systematic error of 10\% for \lir\ and include this in the quoted $1\sigma$ uncertainty, all calculations, and on all figures. Appendix Section \ref{pahmodel} describes our procedure for calculating PAH line luminosities and $z_{\mathrm{\tiny IRS}}$.}
    \tablenotetext{a}{See \cite{Kirkpatrick2012} for details on stellar mass calculations.}
    \tablenotetext{b}{\fagn\ is the integrated AGN power-law emission divided by the total mid-IR IRS flux, and is calculated using the mid-IR decomposition technique of \cite{Pope2008} and \cite{Kirkpatrick2015}. We re-fit this template-based model using MCMC and and calculate \fagn\ at each step in the Markov chain. Tabulated values for \fagn\ correspond to the mean of this distribution. Given the data in hand, \fagn\ can be measured with an accuracy of $\pm\,0.1$ (e.g., \citealt{Pope2008,Kirkpatrick2015}).  }
    \tablenotetext{c}{The 11.3 $\mu$m PAH feature is redshifted out of GS IRS46's \textit{IRS} spectrum.}
    \tablenotetext{d}{GS IRS46 is outside of GOODS-S and CANDELS, preventing the calculation of a stellar mass with comparable methods to the rest of the sample for which deeper data is available. }
\end{deluxetable*}


PDR densities are much greater than the critical density of \cii\ with its primary collisional partners H and H$_2$ ($n_{\mathrm{\tiny crit,\ H}}=3000$ cm$^{-3}$, $n_{\mathrm{\tiny crit,\ H_2}}=6100$ cm$^{-3}$), both of which are heated by  photoelectrons from PAH grains \citep{TH1985,Wolfire1990,Kaufman1999,Malhotra2001,Goldsmith2012}. Thus, a correlation between \cii\ and PAH emission is likely if both lines originate from the same PDR regions. Indeed, \cite{Helou2001} found the ratio of \cii\ emission over integrated $5-10$ $\mu$m flux in star-forming galaxies to be independent of far-IR color, which strongly favors a co-spatial origin.

\cite{Pope2013} report a deficit of 6.2 $\mu$m PAH emission at higher \lir\ in (ultra) luminous IR galaxies (LIRGs: \loglir$=11-12$, ULIRGs: \loglir$>12$) and sub-millimeter (mm) galaxies, a feature also observed for \cii\ emission in similar galaxy populations. Indeed, the luminosity ratio of \cii\ to \lir\ decreases at higher \lir\ in local and high-$z$ galaxies.\footnote{E.g., \citealt{Malhotra1997,Malhotra2001,Luhman1998,Luhman2003,Helou2001,DiazSantos2013,DiazSantos2014,DiazSantos2017,Stacey2010,Magdis2014,Rigopoulou2014,Brisbin2015,Ibar2015,Zanella2018,Rybak2019}.} In low-$z$ (U)LIRGs, \cite{DiazSantos2013,DiazSantos2017} find \lcii/\lir\ empirically anti-correlates with average dust temperatures and IR luminosity surface densities, suggesting that either harder and more intense radiation fields lower the \lcii/\lir\ ratio, or larger dust grains out-compete PAHs for ionizing photons, starving the gas. \cite{Smith2017} find the star-formation rate surface density to be a primary factor driving the \cii-deficit, reconciling nearby resolved measurements and high$-z$ galaxies with a relation that spans over six orders of magnitude. At $z\sim3$, \cite{Rybak2019} find evidence for thermal saturation of C$^+$ as the primary driver of the deficit (see also \citealt{MunozOh2016}). Other potential contributors to the \cii-deficit include positive PAH grain charging where fewer photoelectrons are available to collisionally excite \cii\ (e.g., \citealt{Helou2001}), density effects (e.g., \citealt{Smith2017}), and/or \cii\ self-absorption, although the latter scenario requires unusually large gas column densities in PDRs and is unlikely \citep{Luhman1998, Malhotra2001}. 

Regardless of its origin, the \cii-deficit implies that one of the most important cooling lines for star-formation falls off in luminosity at higher \lir, or equivalently, higher star-formation rate (SFR, \citealt{Kennicutt1998}). This implies a change in one or all of the following: the photoelectric heating efficiency of the ISM, far-UV radiation field strength and hardness, gas density and PDR geometry \citep{Smith2017}. Furthermore, galaxies have higher SFR per unit stellar mass at earlier times than they do locally (e.g., \citealt{MadauDickinson2014}), suggesting that ISM conditions evolve as a function of redshift and SFR \citep{Scoville2017,Tacconi2018,Liu2019}. Indeed, \cite{Stacey2010} found that the \cii-deficit is pushed to higher \lir\ at higher redshifts; however, \cite{Zanella2018} did not observe this offset in a sample of main-sequence galaxies at $z\sim2$. In either case, all galaxies may follow the same \lcii/\lir\ trend as a function of \lir\ normalized by molecular gas mass \citep{Stacey2010,GC2011}. Thus, the gas cooling properties and stellar radiation field strengths in local and $z>1$ star-forming galaxies could be comparable for a given star-formation efficiency (SFE$\,\equiv\,$SFR$\,/\,$M$_{\mathrm{H}_2}$). If this is the case, high$-z$ star-formation could be a scaled up version of star-formation today with comparable ISM conditions, and therefore, similar mid- and far-IR PDR line ratios. 

In this paper, we combine new observations using the Atacama Large Millimeter/submillimeter Array (ALMA) to investigate the properties of ISM heating and cooling in $z\sim2$ star-forming galaxies through combined observations of \cii\ and PAH emission. With archival \textit{Spitzer Infrared Spectrograph} (\textit{IRS}) spectra, we can identify pure star-forming galaxies to study the properties of $z\sim2$ PDRs and star-formation without concern for feedback from an active galactic nucleus (AGN). Using ratios of \cii\ to PAH emission, we investigate the photoelectric efficiency in PDRs near the peak in the Universe's star-formation rate density, a critical epoch for galaxy evolution during which most of the stellar mass in the present day Universe was assembled (e.g., \citealt{MadauDickinson2014}). We investigate the evolution in \cii/PAH emission with redshift, and comment on the technical aspects of synergistic surveys combining ALMA and mid-IR spectrographs, with applications to the \textit{James Webb Space Telescope} Mid-Infrared Instrument (\textit{JWST}/MIRI). 

The paper is organized as follows: In Section 2 we present the galaxy sample, selection criterion, and observations including novel and archival data. Our analysis techniques and emission line measurements are described in Section 3. We present our results in Section 4 and discuss their implications in Section 5. Section 6 summarizes our conclusions. Throughout this work we assume a Salpeter IMF and adopt a $\Lambda$CDM cosmology with $\Omega_m=0.3$, $\Omega_\Lambda=0.7$, and $H_0=70$ km s$^{-1}$ Mpc$^{-1}$. 


\begin{figure}[t!]
    \includegraphics[width=0.48\textwidth]{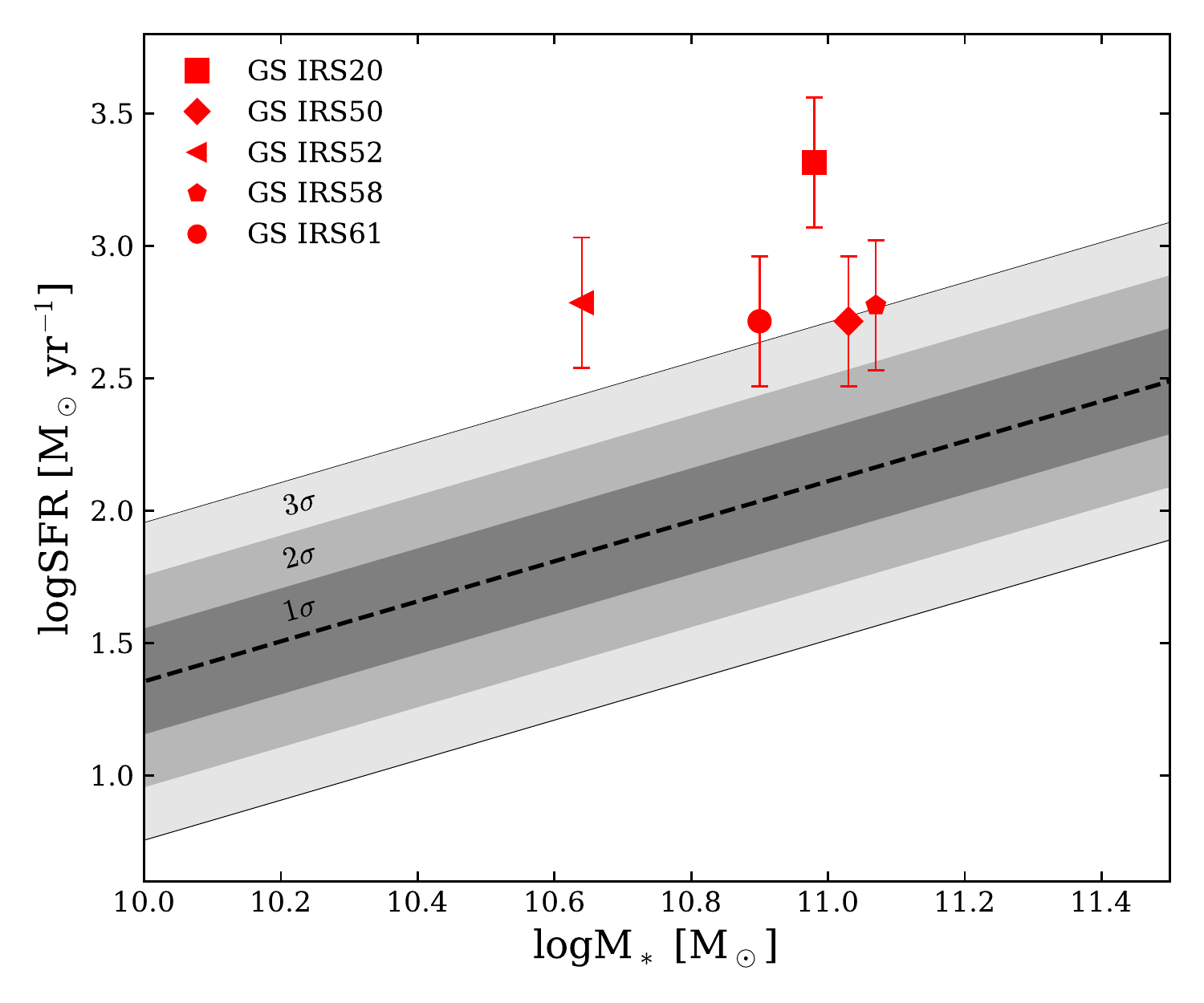}
    \caption{The $z\sim2$ star-forming main-sequence of \cite{Speagle2014} (dashed black line) assuming the same (Salpeter) IMF used in our calculations. We shade 1, 2, and $3\sigma$ intrinsic scatter about the main-sequence in gray. Galaxies from our sample are shown in red, excluding GS IRS46 which does not have a robust stellar mass estimate and is not detected in \cii\ with ALMA. We only detect \cii\ emission in GS IRS20 (red square), the starburst galaxy located $>5\sigma$ above the main sequence.  \label{sfms}}
\end{figure}


\begin{figure*}
    \gridline{\fig{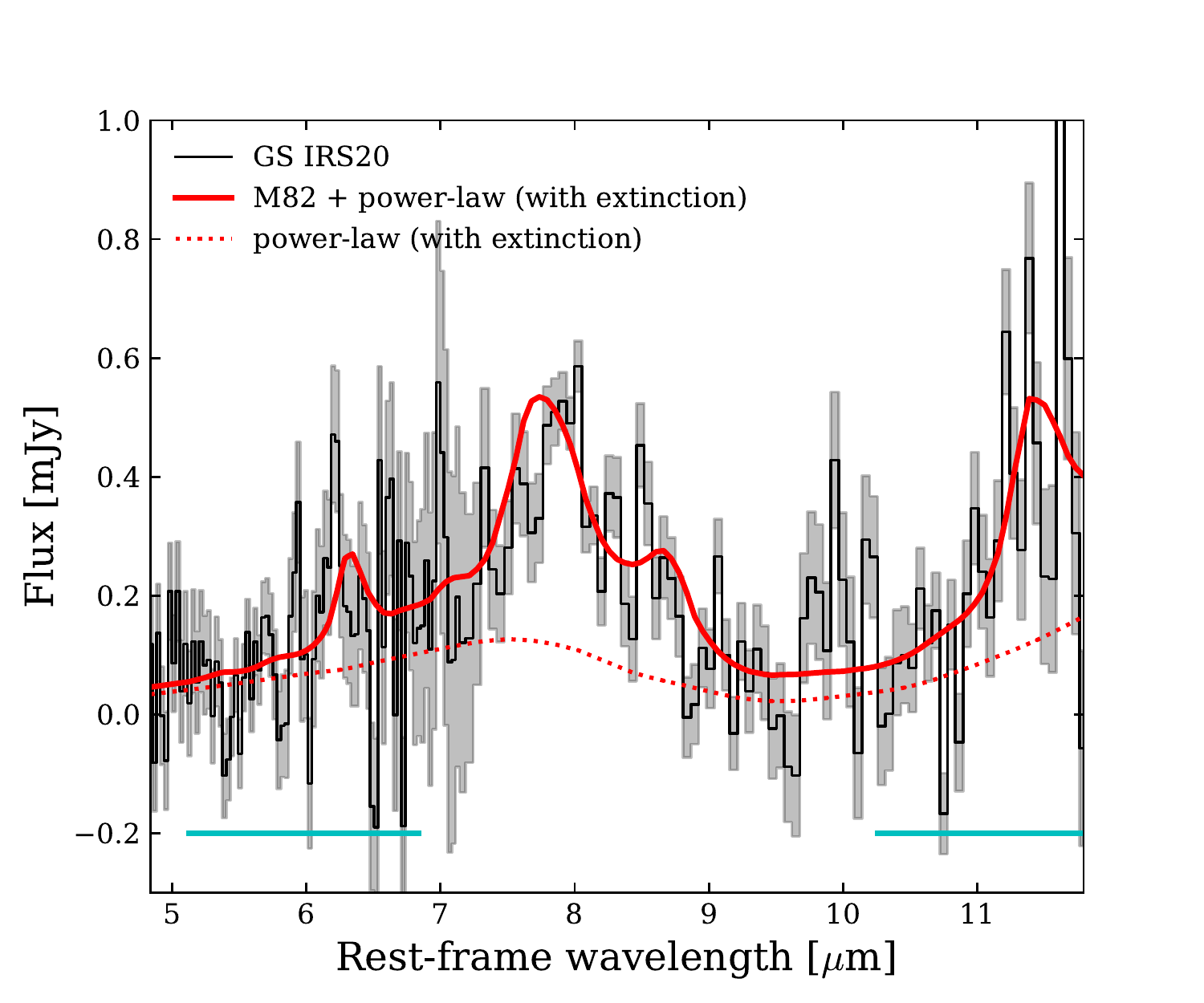}{0.33\textwidth}{}
          \fig{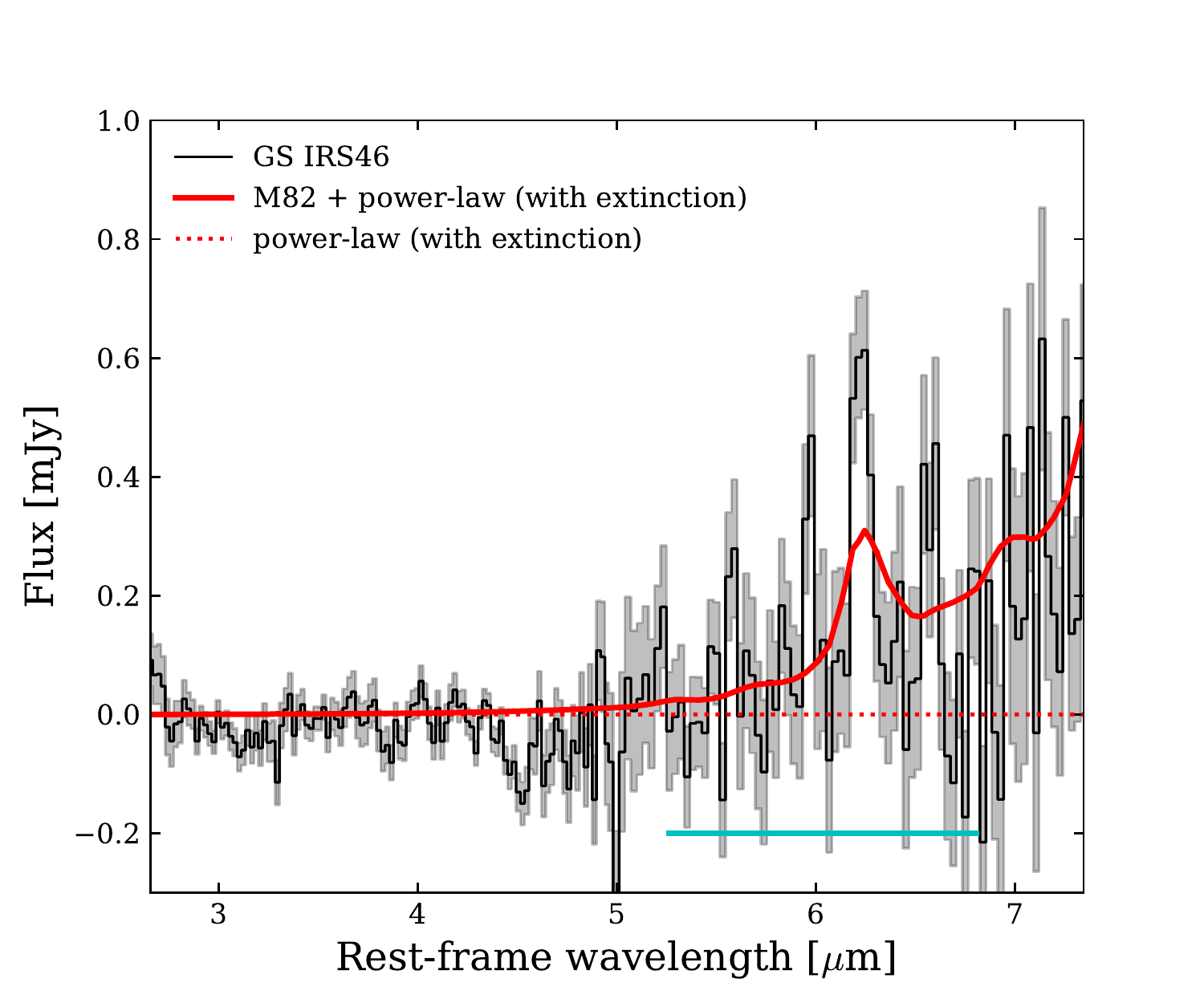}{0.33\textwidth}{}
          \fig{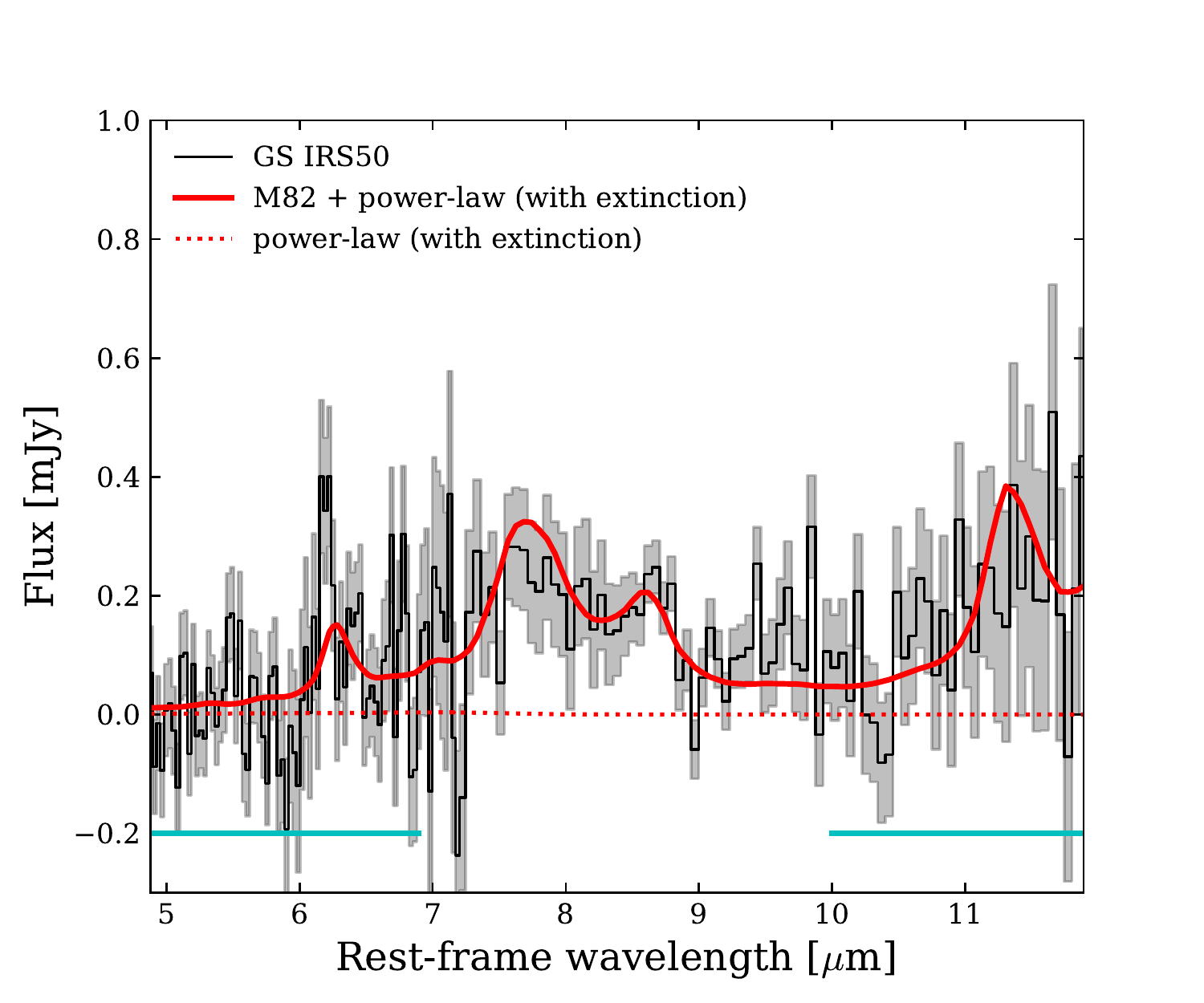}{0.33\textwidth}{}}
    \vspace{-3em}
    \gridline{\fig{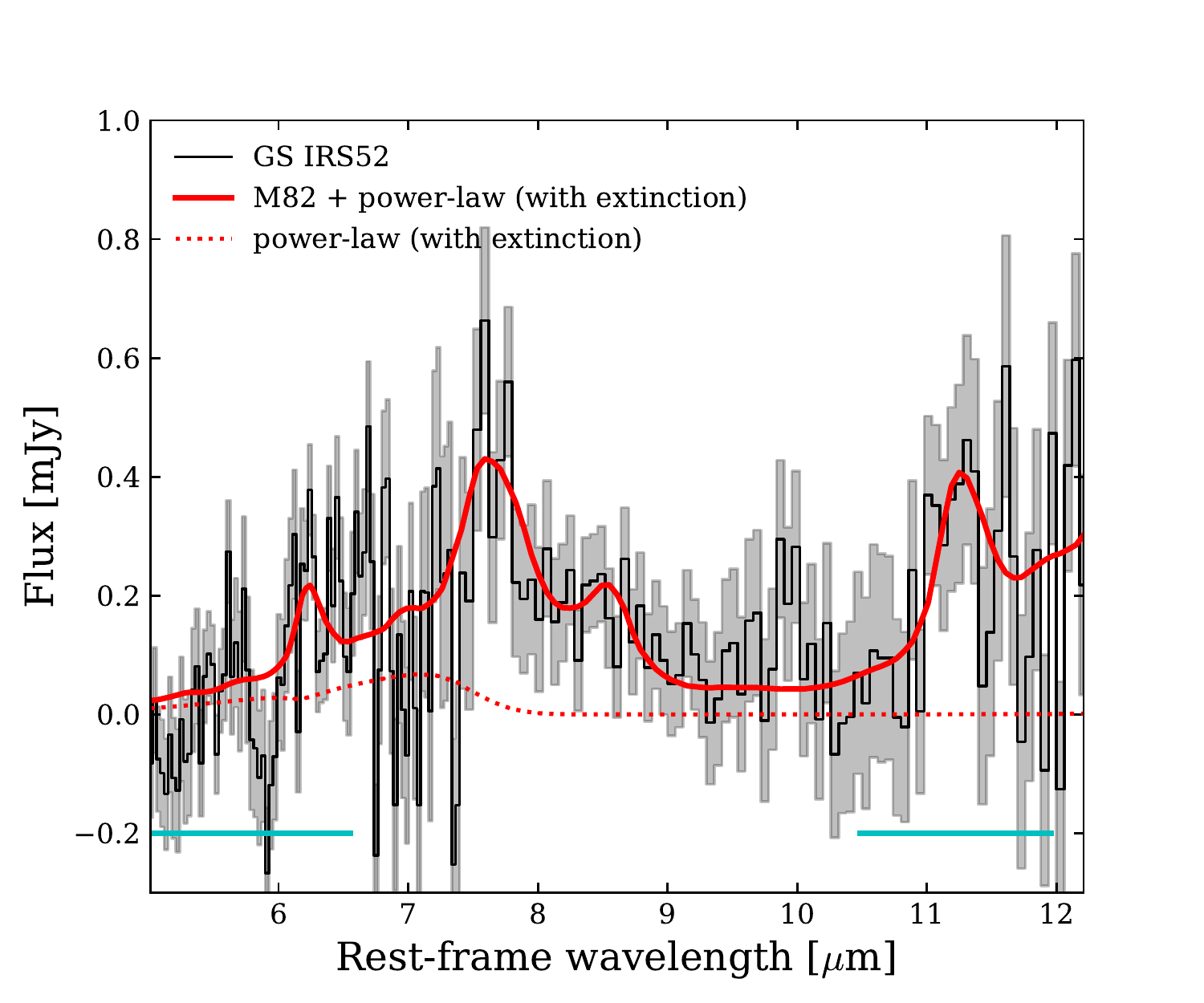}{0.33\textwidth}{}
          \fig{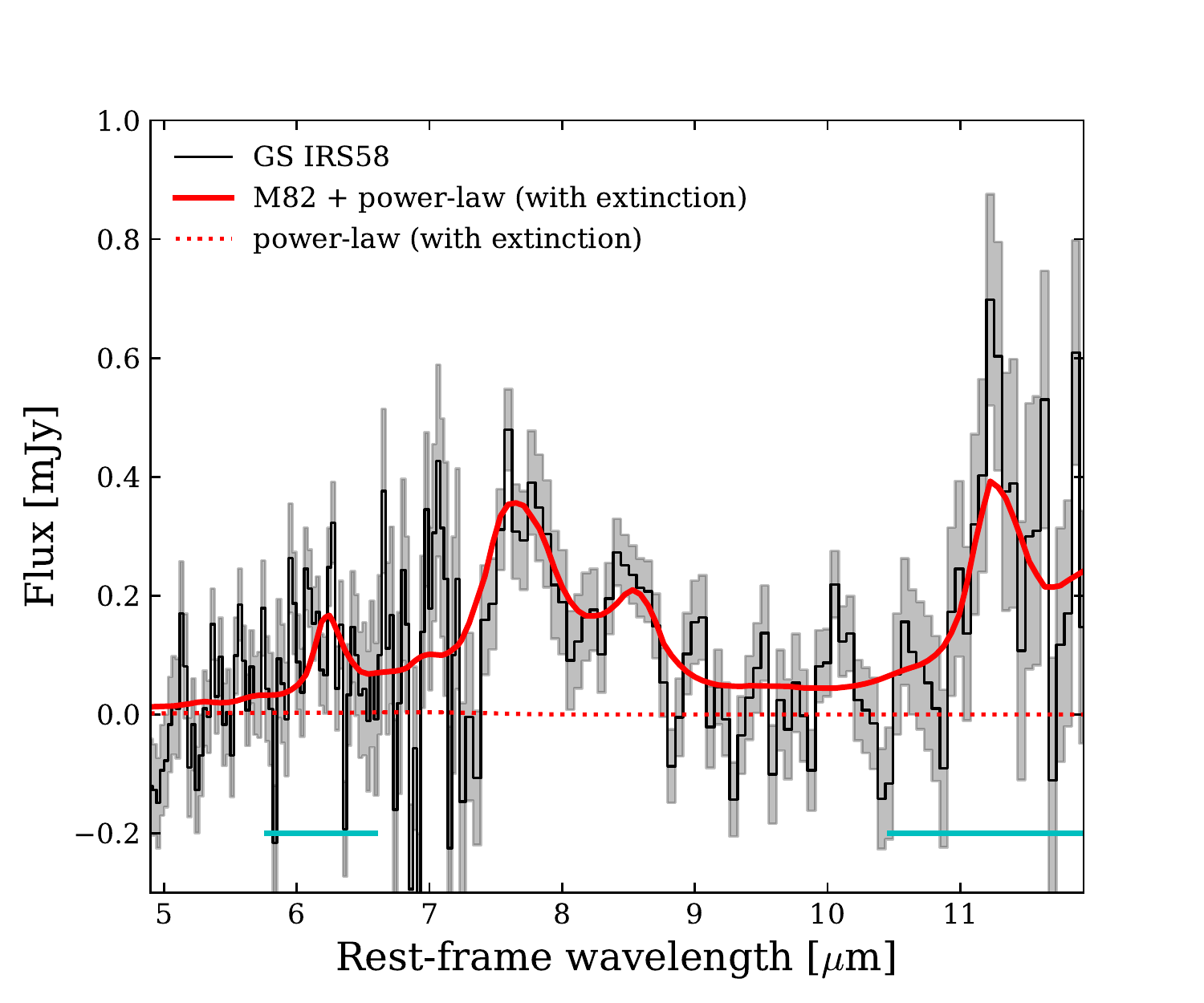}{0.33\textwidth}{}
          \fig{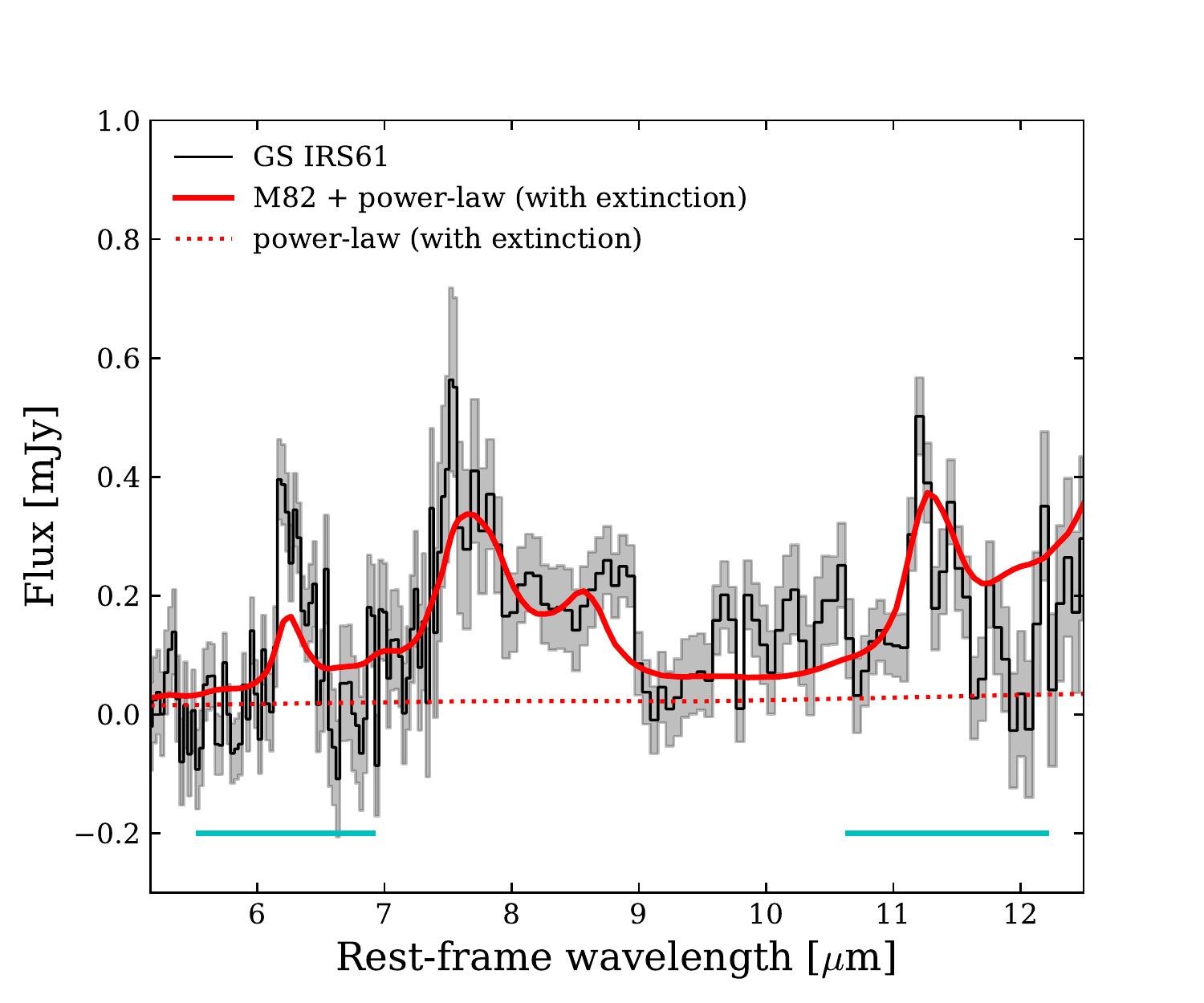}{0.33\textwidth}{}}
    \vspace{-2em}
    \caption{\textit{Spitzer IRS} spectra for the 6 targets observed with ALMA. The \textit{IRS} spectra are shown in black, with uncertainties shaded in gray. The red solid line corresponds to best-fit empirical M82 templates on top of a power-law continuum component (red dotted line). The simple fit is used too quantify the AGN fraction in the mid-IR and we employ a more sophisticated model to measure individual line luminosities and redshifts (see Appendix \ref{pahmodel}). Horizontal cyan lines show the regions where we fit Lorentzian profiles to the 6.2 $\mu$m and 11.3 $\mu$m PAH features.\label{irsSpectra}}
\end{figure*}


\section{Sample and Observations}
\subsection{Sample Selection}
We have assembled a sample of six IR-luminous galaxies (\loglir$>12$) at $z=1.7-2$ with extensive coverage from restframe ultraviolet to sub-mm wavelengths, selected primarily by the presence of luminous PAH features in the mid-IR and little to no underlying power-law continuum. These systems are dominated by star-formation: an AGN would heat dust to high temperatures and emits warm black-body emission at mid-IR wavelengths which we do not detect (e.g., \citealt{Laurent2000,Sturm2000,Tran2001,Sajina2007}). Our sample comes from a larger multiwavelength parent catalog described in \cite{Kirkpatrick2015}. To summarize, multiwavelength data was collected for 343 (U)LIRGs between $z=0.3-2.8$ in the Great Observatories Origins Deep Survey North/South (GOODS-N/S), Extended \textit{Chandra} Deep Field Survey (ECDFS), and the \textit{Spitzer} Extragalactic First Look Survey fields. The primary target selection criterion was the presence of mid-IR spectroscopy from \textit{Spitzer IRS}. For more details on the parent sample selection method, we refer readers to Section 2.1 of \cite{Kirkpatrick2015}. 

With our ALMA cycle 5 program targeting \cii\ emission at $z\sim2$, we observed six star-forming galaxies between $z=1.7-1.9$ from the \cite{Kirkpatrick2015} sample with L$_{\mathrm{\tiny PAH,}6.2}$/\lir$>0.004$ and \loglir$>12$. These galaxies all have little to no evidence of AGN contamination to the mid-IR spectrum (\fagn), based on \textit{IRS} spectral decomposition, and as evidenced by their 6.2$\mu$m PAH equivalent widths \ewPAH$>0.5\,\mu$m, which is the threshold established in nearby (U)LIRGs for star-formation dominated systems \citep{Stierwalt2014}. The selection of sources based on strong PAH features in high IR-luminosity galaxies has been shown in the literature to be a robust way for selecting galaxies with minimal AGN contamination (e.g., \citealt{Houck2005,Yan2005,Brandl2006,Sajina2007,Smith2007,Armus2007,Pope2008,Veilleux2009,Kirkpatrick2012}). 

Configuring the ALMA Band 9 Local Oscillator to efficiently target \cii\ over the redshift range spanned by our sample was a challenging factor in the design of our experiment. Efficient programs capable of observing multiple targets with minimal baseband tunings are optimal for taking advantage of limited high-frequency ALMA observing time. To maximize sample size while minimizing overhead, we manually configured each spectral window within the Band 9 constraints to cover \cii\ in multiple galaxies in a given ALMA Science Goal. 

Most of the galaxies in our sample have robust stellar masses constrained by deep \textit{HST} and \textit{Spitzer} photometry. Galaxies in our sample are high mass, LogM$_*/$M$_\odot=10.6-11$, and dusty, as evidenced by \textit{Spitzer} and \textit{Hershel} photometry. Figure \ref{sfms} shows the star-forming main-sequence (MS) of galaxies at $z\sim2$ taken from \cite{Speagle2014}. Our sample lies above the $z\sim2$ SFMS, with $\log\Delta$SFMS (the observed SFR over the MS SFR for the same stellar mass) between $0.6-1$ dex for most galaxies in our sample and as high as $\sim1.2$ dex in GS IRS20, well within the starburst domain. Table \ref{sampsum} summarizes global properties for galaxies in our sample. 


\subsection{Multiwavelength Observations\label{sec:2.2}}
Our sources are in ECDFS and were selected to have mid-IR spectroscopy from the \textit{Spitzer} \textit{IRS} \citep{Fadda2010,Kirkpatrick2012,Kirkpatrick2015}. A full description of \textit{IRS} observations, data reduction, and sample-selection can be found in \cite{Pope2008} and \cite{Kirkpatrick2012}. The extracted spectra are shown in Figure \ref{irsSpectra} with simple fits to the mid-IR emission that we use to calculate \fagn; more sophisticated model fits are employed to measure PAH line luminosities (see Appendix \ref{pahmodel}). In addition to \textit{Spitzer} IRS spectra, photometry from \textit{Herschel} (PACS and SPIRE), and \textit{Spitzer} (IRAC and MIPS) is available for all targets (see \citealt{Kirkpatrick2015} for details).


ECDFS includes the GOODS-S field, which was covered by the Cosmic Assembly Near-IR Deep Extragalactic Legacy Survey (CANDELS; \citealt{Grogin2011,Koekemoer2011}), providing deep WFC3/IR imaging for five out of six galaxies in our sample. We downloaded the \hband\ and $Z_{850lp}$ field maps, and correct for the known systematic astrometric offset of $0.08''$ in RA and $0.26''$ in DEC relative to ALMA's astrometry in GOODS-S \citep{Elbaz2018}. Thumbnail images for our sample are shown in Figure \ref{imgs}. We also matched our galaxies to visual morphological classifications presented in \cite{Kartaltepe2015} to assess the incidence of mergers in the sample. 


\begin{figure*}
    \gridline{\hspace{-1.2em}
          \fig{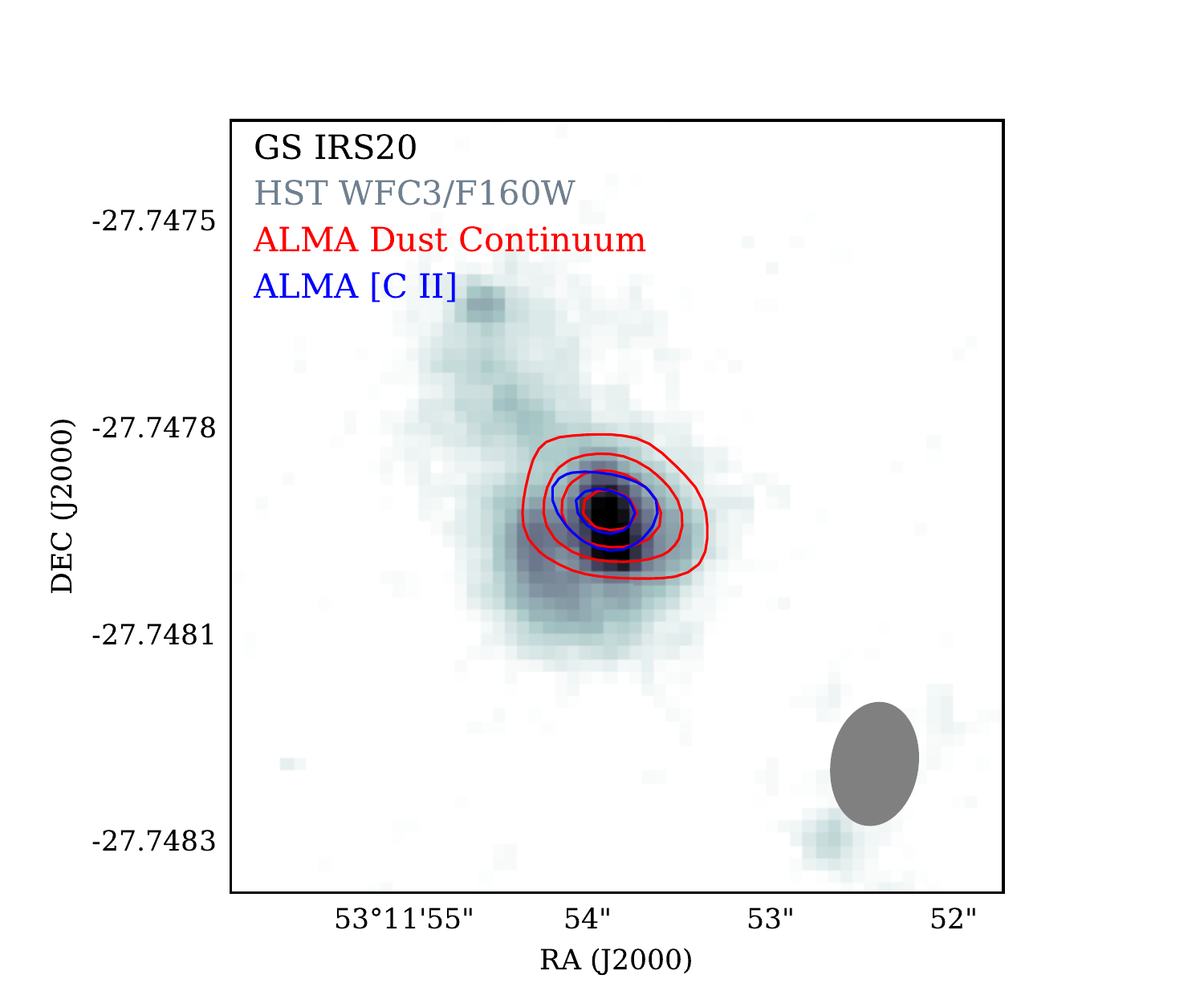}{0.4\textwidth}{}\hspace{-3.5em}
          \fig{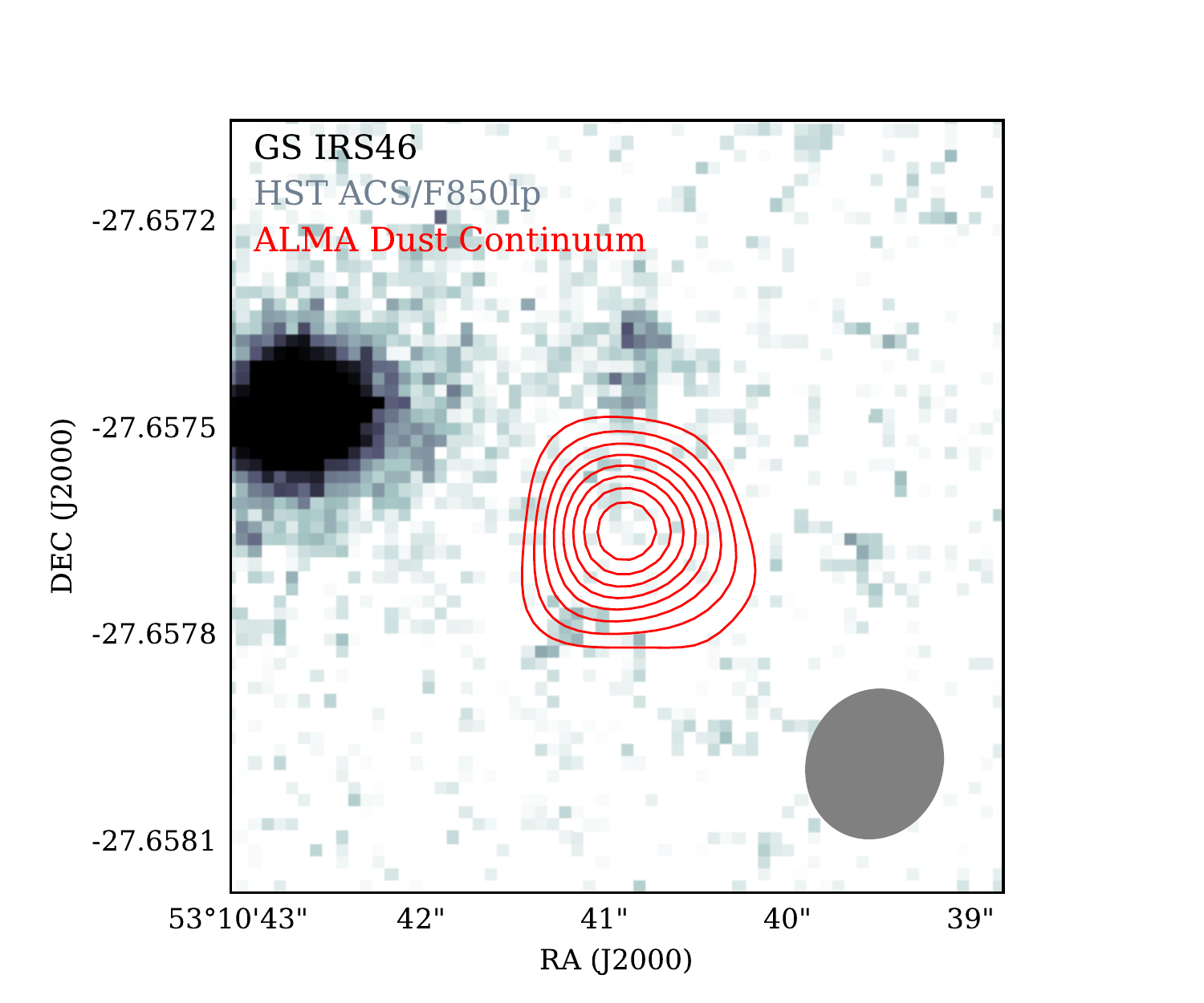}{0.4\textwidth}{}\hspace{-3.5em}
          \fig{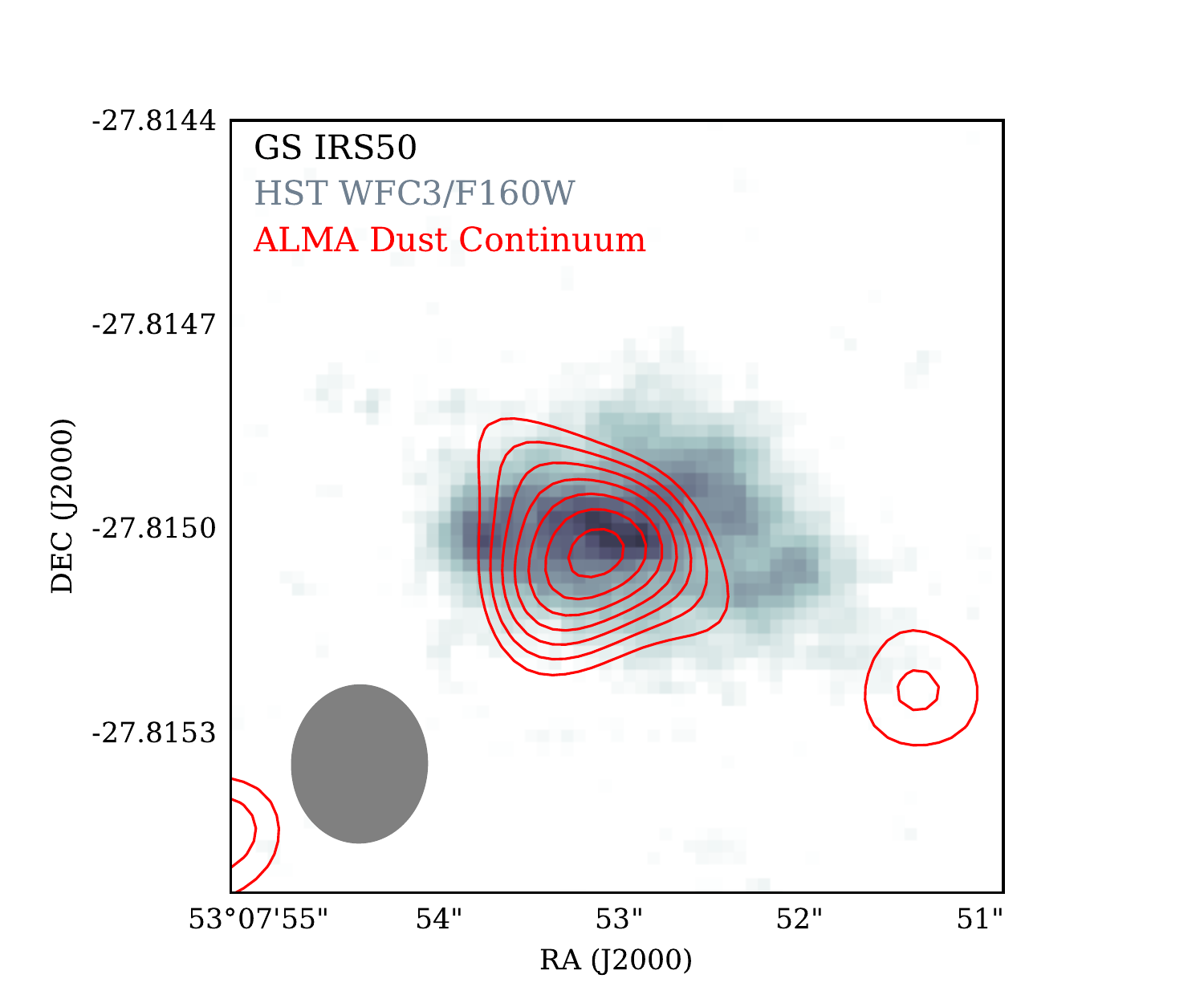}{0.4\textwidth}{}}
    \vspace{-3em}
    \gridline{\hspace{-1.2em}
          \fig{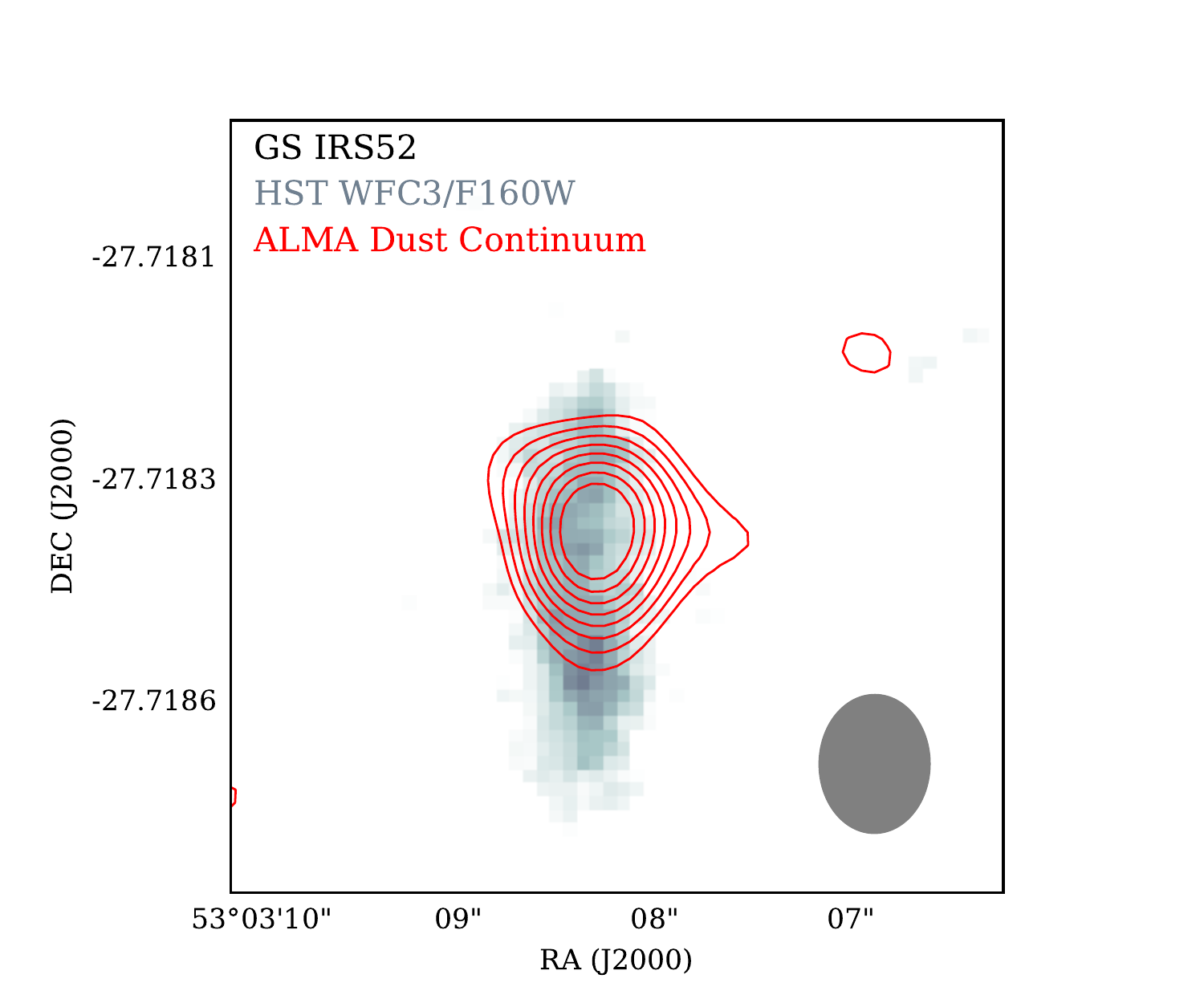}{0.4\textwidth}{}\hspace{-3.5em}
          \fig{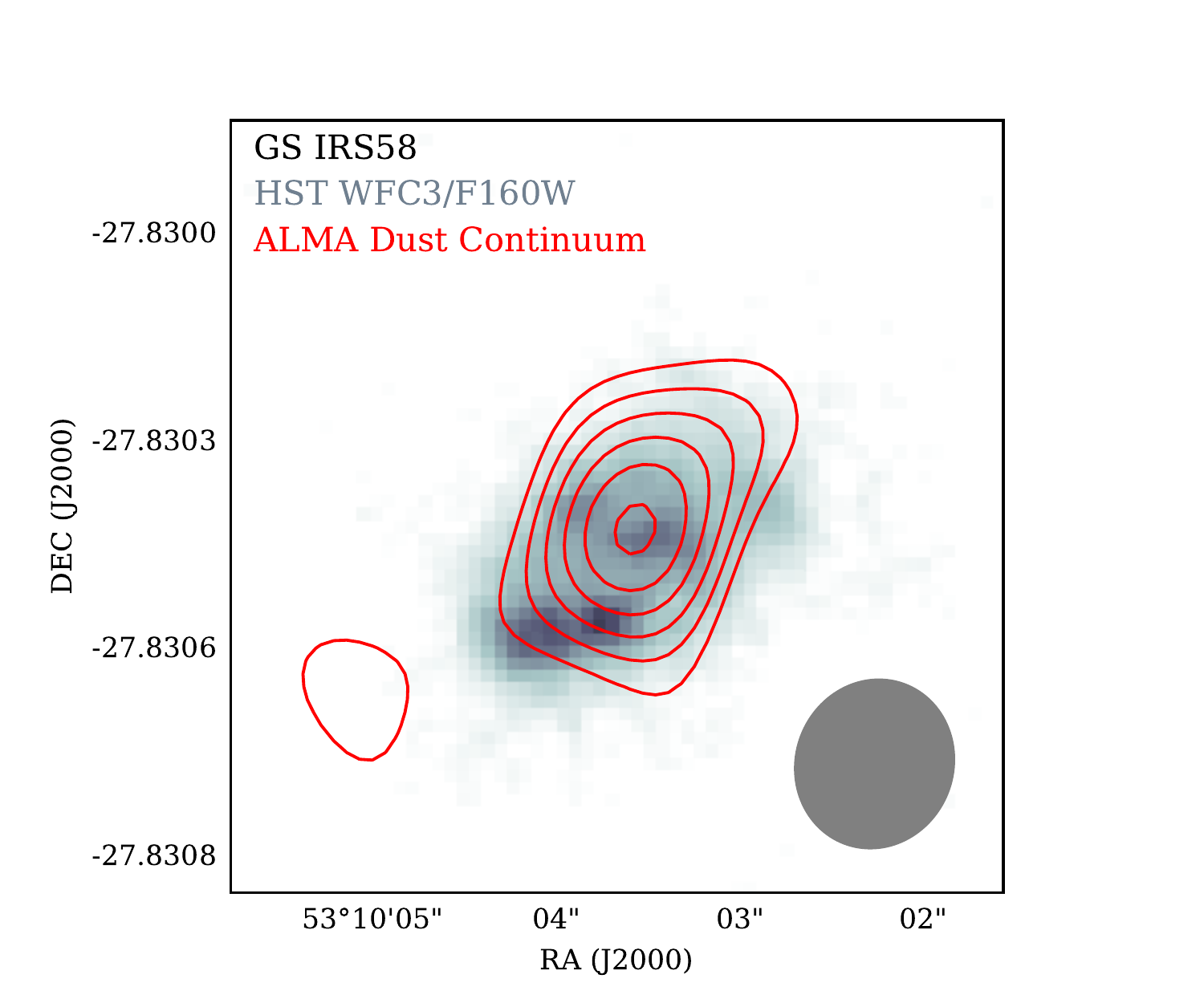}{0.4\textwidth}{}\hspace{-3.5em}
          \fig{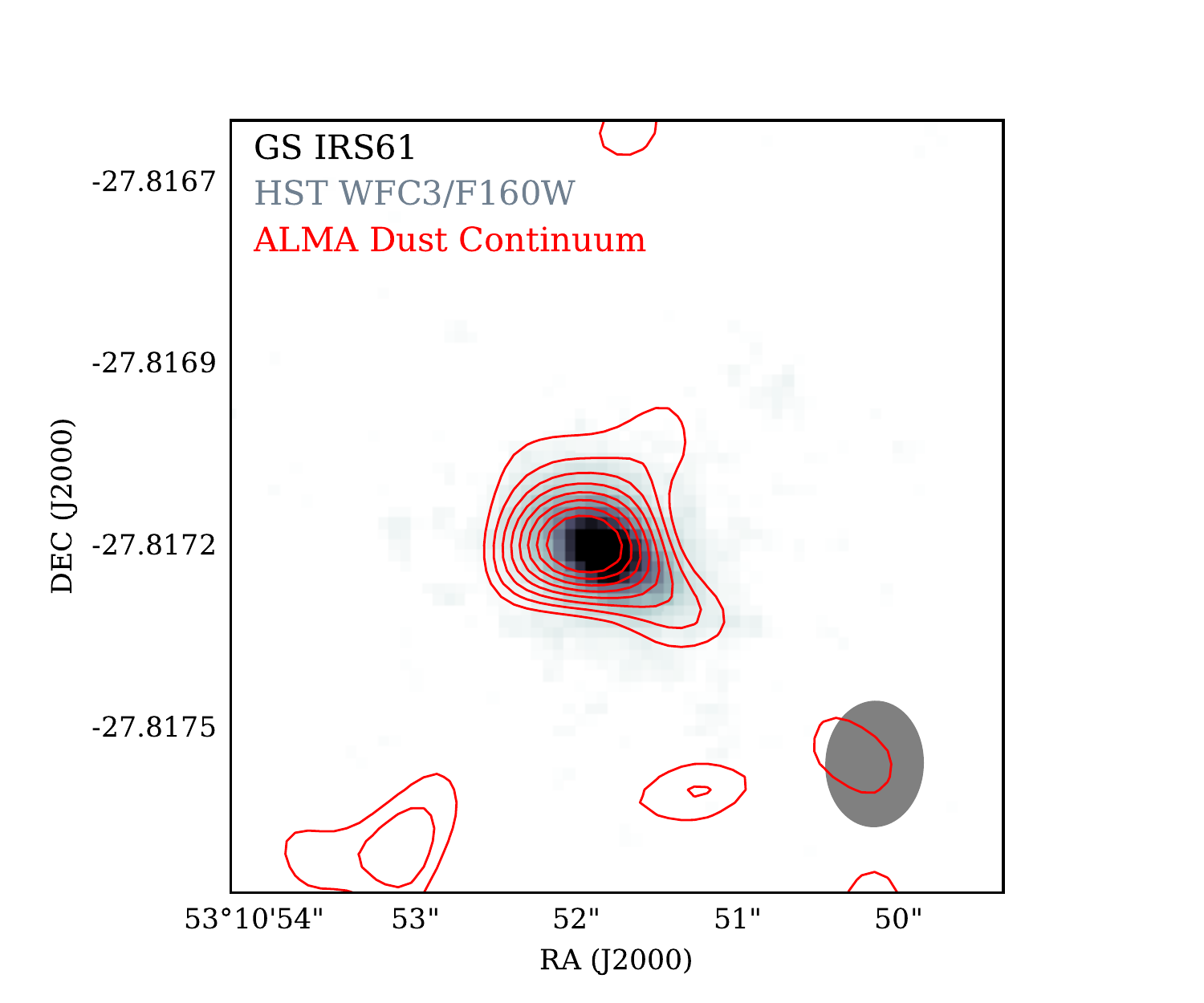}{0.4\textwidth}{}}
    \vspace{-2em}
    \caption{Postage stamp images for each of the targeted galaxies. \textit{HST}/WFC3 \hband\ imaging is shown in the background when available, and ALMA dust continuum contours are overplotted in red. For GS IRS46, we show \textit{HST}/ACS F850lp (z-band) maps in the background \citep{Giavalisco2004}. In the case of GS IRS20, we also show integrated \cii\ emission contours in blue. The ALMA beam is represented by a gray ellipse in all images. \label{imgs}}
\end{figure*}


\subsection{ALMA Observations and Data Processing \label{almaobs}}

We carried out ALMA Band 9 observations of our targets during Cycle 5 (PI A. Pope, Project ID: 2017.1.01347.S) targeting \cii\ emission at restframe $157.74\ \mu$m. For the range of redshifts in our sample, \cii\ is redshifted to an observed frequency of $653.36-686.38$ GHz. We estimated integration times necessary to detect the \cii\ line at $10\sigma$ for galaxies in our sample by assuming a conservative \lcii/\lir\ ratio of 0.002 and \cii\ line width of 300 km s$^{-1}$, characteristic of existing \cii\ detections $z\sim2$ galaxies prior to our observations \citep{Stacey2010}. The minimum predicted \cii\ flux for all galaxies in the sample was 15 Jy km s$^{-1}$, which we used to set the integration time for each observation by requiring a $>10\sigma$ line detection, or equivalently, a sensitivity of 5 mJy over 300 km s$^{-1}$ bandwidth. 

To avoid resolving out \cii\ emission at $z\sim2$, we requested an angular resolution of $\sim0.5''$. The observations took place in July 2018 in ALMA configuration C43-1 which has an angular resolution of $0.52''$ at 650 GHz and maximum recoverable scale of $4.4''$, corresponding to $36.5$ kpc at $z=2$. The expected radii of sub-mm and \textit{HST} \hband\ emission in $z\sim2$ star-forming galaxies is $<8$ kpc (e.g., Fig. \ref{imgs}, \citealt{Zanella2018,CalistroRivera2018,Lang2019}), so it is unlikely that our observations are missing flux on large scales due to interferometric spatial filtering. Six galaxies in our proposal were observed for $\sim18$ minutes on-source, achieving the target sensitivity of 5 mJy over 300 km s$^{-1}$ bandwidth at a native resolution of 31.250 MHz (13.6 km s$^{-1}$) which was later re-binned to lower spectral resolutions.  

The data were reduced using the standard ALMA pipeline in CASA v5.1.1-5 \citep{CASA}. We first imaged the data using \texttt{tclean} with Briggs weighting in continuum-mode, iteratively adjusting the robust parameter $R$ to maximize the ratio of peak continuum emission to map RMS. We extracted peak and integrated continuum flux densities through elliptical apertures which were set by fitting a 2D Gaussian function to the bright continuum emission in each observation. We detect continuum emission at representative frequencies of $652-699$ GHz in all of our targets at signal-to-noise (SNR) between $7.5-17.8$. After verifying the presence of underlying continuum, we created a linear continuum model in the $uv$-plane, taking care to mask out high-amplitude visibilities that could correspond to potential line emission. Next, we continuum-subtracted the ALMA cubes in the $uv$-plane and imaged the spectral windows with \texttt{tclean} and Briggs weighting using $R=0.5$. Final continuum measurements, ALMA beam characteristics, and spectral line statistics are given in Table \ref{almacont}. 


\renewcommand{\tabcolsep}{4pt}
\begin{deluxetable*}{lccccccccc}
    \tablecaption{ALMA Cycle 5 Band 9 Observations: Continuum Imaging and Spectral Line Data\label{almacont}}
    \tabletypesize{\footnotesize}
    \tablehead{ & &  & & Continuum Maps & &  \\\cmidrule{3-7} Target & Beam FWHM & $\lambda_{\tiny\mathrm{obs}}$\tablenotemark{a} & rms  & Peak Flux   &  Integrated Flux & \Reff & $p(l|\mbox{ALMA,}\Delta z)$\tablenotemark{b} & F$_{\mathrm{\tiny [C II]}}$\tablenotemark{c} \\ & [arcsec] & [$\mu$m] &  [mJy/beam] &  [mJy/beam] &  [mJy/beam] & [kpc] &  & [mJy]}
    \startdata 
    GS IRS20 & 0.51 $\times$ 0.37 & 466.60 & 0.87 & $14.38\pm0.87$ & $17.3\pm1.1$  & 1.81 & 0.68   & $331.59\pm9.66$   \\[.3ex]
    GS IRS46 & 0.74 $\times$ 0.66 & 456.95 & 0.54 & $6.55\pm0.60$  & $9.3\pm1.0$   & 2.63 & 0.56   & ($<10.5$) \\  [.3ex]
    GS IRS50 & 0.78 $\times$ 0.67 & 465.07 & 0.41 & $4.82\pm0.67$  & $7.1\pm1.2$   & 2.72 & 0.29   & ($<14.0$) \\[.3ex]
    GS IRS52 & 0.63 $\times$ 0.51 & 444.77 & 0.10 & $5.21\pm0.51$  & $8.64\pm0.92$ & 2.15 & 0.01   & ($<4.3$ )\\[.3ex]
    GS IRS58 & 0.93 $\times$ 0.86 & 456.95 & 0.13 & $5.49\pm0.70$  & $10.5\pm2.4$  & 3.36 & 0.22   & ($<5.3$)  \\[.3ex]
    GS IRS61 & 0.70 $\times$ 0.54 & 441.62 & 0.95 & $3.70\pm0.35$  & $4.22\pm0.64$ & 2.34 & 0.93   & $<4.6$ \\[.3ex]
    \enddata
    \tablenotetext{a}{ Effective wavelength of collapsed ALMA Band 9 cube.}
    \tablenotetext{b}{ The probability of observing the target's redshifted \cii\ line given all redshift uncertainty and the ALMA Band 9 spectral window configuration. See Equation \ref{pz} in Section 3.3.}
    \tablenotetext{c}{ \cii\ line flux. Upper limits are $3\sigma$ and given for each galaxy, although values in parenthesis are considered unreliable given the low probability of having observed the line. }
\end{deluxetable*}


\section{Analysis}
\subsection{\cii\ Detection in GS IRS20}

Whereas the dust continuum is clearly detected with ALMA for all 6 galaxies (red contours in Fig. \ref{imgs}), \cii\ 158$\mu$m emission is clearly detected in one of six galaxies in the sample, GS  IRS20, at an observed frequency of 650.2505 GHz. This corresponds to a redshift of \zcii$=1.9239\pm0.0002$, in excellent agreement with the PAH-derived redshift: \zcii$-$\zirs$=0.001$. 

We imaged the cube in 30 \kms\ bins, and extracted a spectrum through an elliptical aperture with FWHM and centroid taken from a 2D Gaussian fit to continuum emission. Figure \ref{spec20} shows the detection of [C II] in GS IRS20's ALMA Band 9 spectrum. Gaps in spectral coverage are the result of limitations when configuring ALMA's spectral windows. We integrated the line over the frequency range where emission rose above the continuum level and measured a flux density of $S_{\mathcii}\Delta v=9.95\pm0.07$ Jy km s$^{-1}$ at a SNR of $34.3$ and line velocity width of $\sim330$ km s$^{-1}$. Next, we calculated the \cii\  line luminosity \lcii\ in solar units following \cite{CarilliWalter2013}:
\begin{equation}
    \mathlcii= 1.04\times10^{-3}\times S_{\mathcii} \Delta v D_L^2 \nu_{obs}\ [\mathrm{L}_\odot]
\label{eqn:lcii}
\end{equation}
where $D_L$ is the luminosity distance in Mpc, and $\nu_{obs}$ is the observed frequency of the line in GHz. From the Band 9 spectrum, we calculate \loglc$/\mathrm{L}_\odot=9.169\pm0.003$ in GS IRS20, the highest SNR detection of \cii\ emission in a $z\sim2$ galaxy to date. From a collapsed ALMA data cube containing only line emission, we find that \cii\ in GS IRS20 is marginally resolved with a spatial FWHM of $0.56''$, corresponding to $\approx4.7$ kpc at $z=1.9239$.}


\subsection{[C II] Line Searches and Upper Limits
\label{linesearch}}

No [C II] emission lines were obvious in the ALMA cubes of GS IRS46, GS IRS50, GS IRS52, GS IRS58, and GS IRS61. To search for marginally detected emission lines, we used a circular aperture with radius $0.5''$ to extract a 50 \kms\ spectrum centered on the source's dust continuum position. Next, we extracted additional spectra through the same circular apertures offset by $0.5''$ from the source's center at various angles, as optical light, dust continuum and \cii\ emission can be spatially offset from one another in high redshift ULIRGs (e.g., \citealt{Zanella2018,CalistroRivera2018}). From the set of extracted spectra, we searched each spectral window for the presence of three channels greater than $2\times$ the local rms. No marginally-significant line emission was discovered in this manner, or in stacks of the extracted spectra. 

Given that 83\% of our observations yielded non-detections, and no data was discarded because of poor atmospheric transmission, two explanations are possible. Either the observations were not deep enough to detect \cii\ and an upper limit may be placed on \lcii; or, the line was missed by our ALMA bandpass tunings. To determine which observations can yield a secure upper limit on \lcii, we calculate $p(l|\mathrm{ALMA},\Delta z)$: the probability our ALMA tunings covered the \cii\ line given all redshift uncertainties and the comparatively narrow bandpass widths. The technique adopted for calculating $p(l|\mathrm{ALMA},\Delta z)$ is described in detail in Appendix Section \ref{appendix:pz}. In summary, we integrate redshift probability distribution functions in spectral domains with ALMA coverage. We found this detailed analysis to be crucial for interpreting the data. Table \ref{almacont} includes values of $p(l|\mathrm{ALMA},\Delta z)$ for all targets.

Amongst the non-detections, only GS IRS61 has $p(l|\mathrm{ALMA},\Delta z)>90\%$. For this galaxy, we first calculate the rms over a spectrum at 50 km s$^{-1}$ resolution (rms$_{50}$), extracted from an aperture centered on the dust continuum. Then, we calculate the $3\sigma$ upper limit on the line luminosity using Equation \ref{eqn:lcii} with $ S_{\mathcii}\Delta v = 3\Delta v(\sqrt{6}\,$rms$_{50}$), assuming $\Delta v=300$ km s$^{-1}$ as is observed in GS IRS20. Our upper limits for GS IRS61 on \lcii\ are summarized in Table \ref{almacont}, and could be a factor 1.8 (0.25 dex) larger than what is reported if we assume a more extreme $\Delta v=600$ km s$^{-1}$, greater than the noise-weighted average of $\sim430$ km s$^{-1}$ as observed in [C II]-emitters at $z\sim2-3$ (e.g., the sample of \citealt{Gullberg2015}). 


\subsection{Morphology}

In all of our observations, dust continuum emission is marginally resolved: the major and minor axes of 2D Gaussian fits to dust emission are equal to $0.5-2''$, slightly greater than the ALMA beam in all cases. We use these size measurements to calculate \Reff, the radius containing 50\% of the total continuum flux at the effective rest-frame wavelengths (approximately  $160\mu$m) of our observations. Table \ref{almacont} includes values of \Reff, which we use to calculate IR surface densities. Given that the extent of dust continuum is marginally greater than the ALMA beam in all cases, our measurements of \Reff\ may be thought of as upper limits. 

Our ability to distinguish substructure in the ALMA maps is limited; however, extended \hband\ emission in the \textit{HST} thumbnails of GS IRS20, GS IRS50 and GS IRS58 suggests disturbed, perhaps merger-driven, morphologies in some cases. We matched our sources to the morphological classification catalog of \cite{Kartaltepe2015} to determine the incidence of mergers in our sample. Each of our targets had the maximum 68 classifications per galaxy. GS IRS20 is considered to be a merger by 80\% of classifiers, and irregular by 100\%, consistent with its position $>5\sigma$ above the $z=2$ galaxy main sequence (Fig. \ref{sfms}), and the presence of faint extended \hband\ emission to the North-East, reminiscent of a tidally disrupted stellar population. Dust continuum and \cii\ emission in GS IRS20 are co-spatial and coincide with the \hband\ maximum. 

The rest of the sample was not classified as mergers, and GS IRS61 is classified as a spheroid by the full set of classifiers. The spatial extent of \hband-band and dust continuum in GS IRS61 is $\sim5$ kpc (FWHM), making this galaxy extremely compact. GS IRS46 is offset from the \textit{HST}/ACS z-band map by $0.70''$ after correcting for the astrometry offset between \textit{HST} and ALMA (see Section \ref{sec:2.2}). This corresponds to $\sim6$ kpc physical offset between the detected stellar light and dust continuum emission in this galaxy. Given the uncertainty introduced by this offset, we do not report a stellar mass or show optical data points for this dusty galaxy. 


\begin{figure}
    \plotone{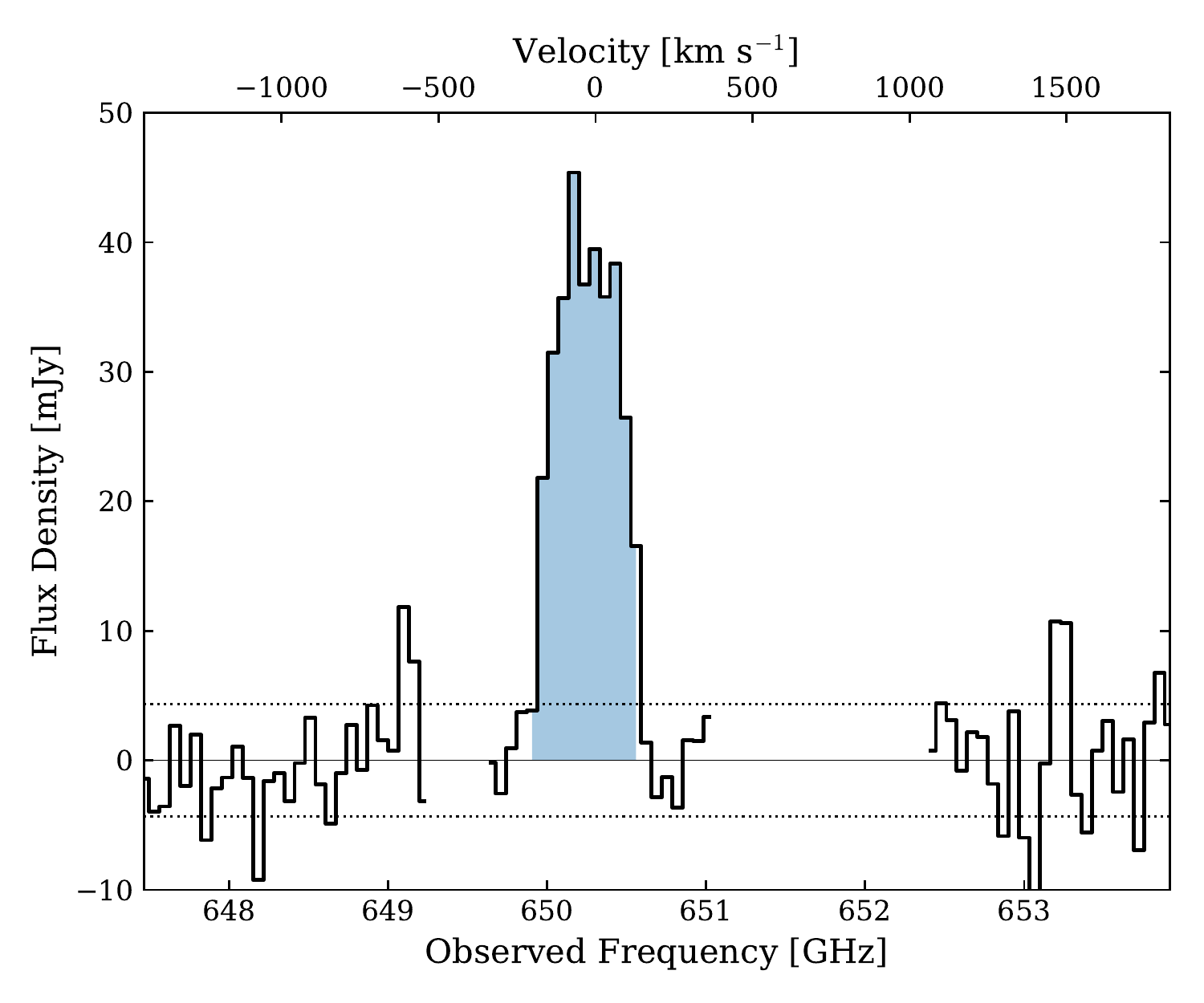}
    \caption{GS IRS20's ALMA Band 9 continuum-subtracted spectrum showing the robust detection of [C II] at $z=1.924$, binned to 30 km s$^{-1}$ resolution. The top axis shows relative velocities in km s$^{-1}$ from the line's centroid. The horizontal dotted lines correspond to $\pm1\sigma$ noise, and the shaded blue region indicates where line emission was integrated: the peak is detected at an SNR of 19.1$\sigma$ and the integrated emission at an SNR of 34.3$\sigma$. Gaps in the spectra are due to the observation's spectral window configuration.   \label{spec20}}
\end{figure}


\begin{figure*}
    \gridline{\fig{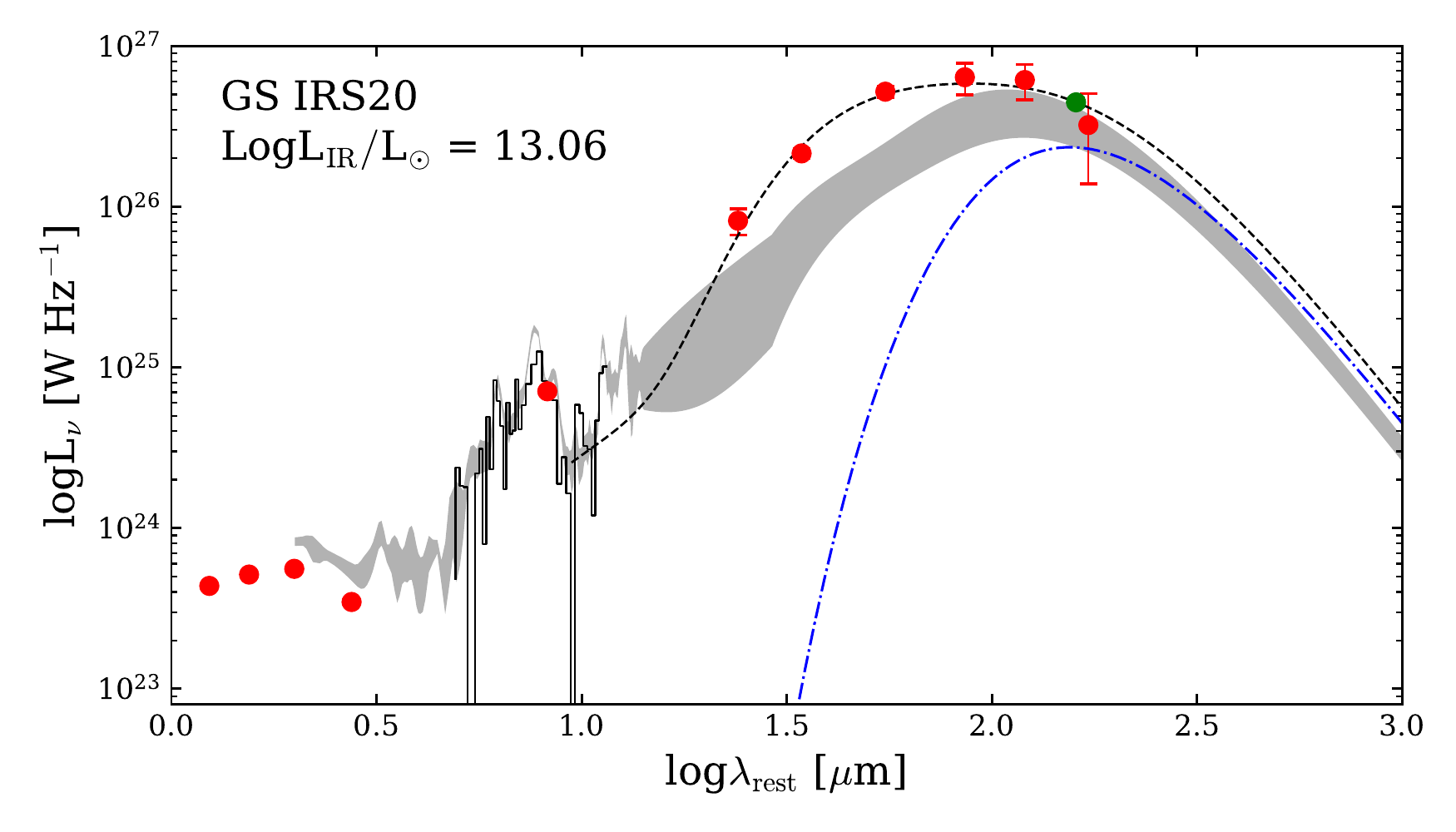}{0.49\textwidth}{}
          \fig{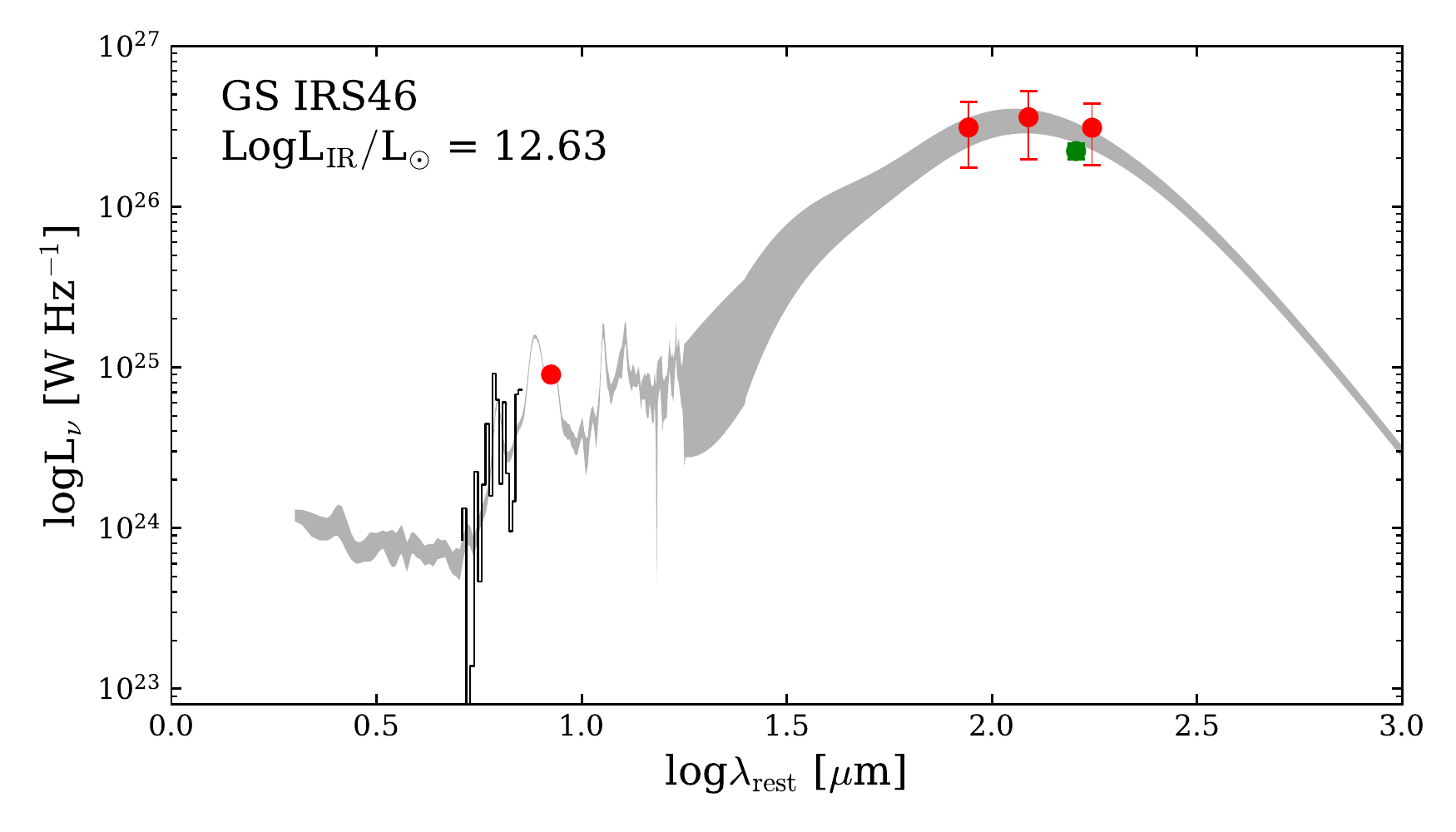}{0.49\textwidth}{}}
    \vspace{-2em}
    \gridline{\fig{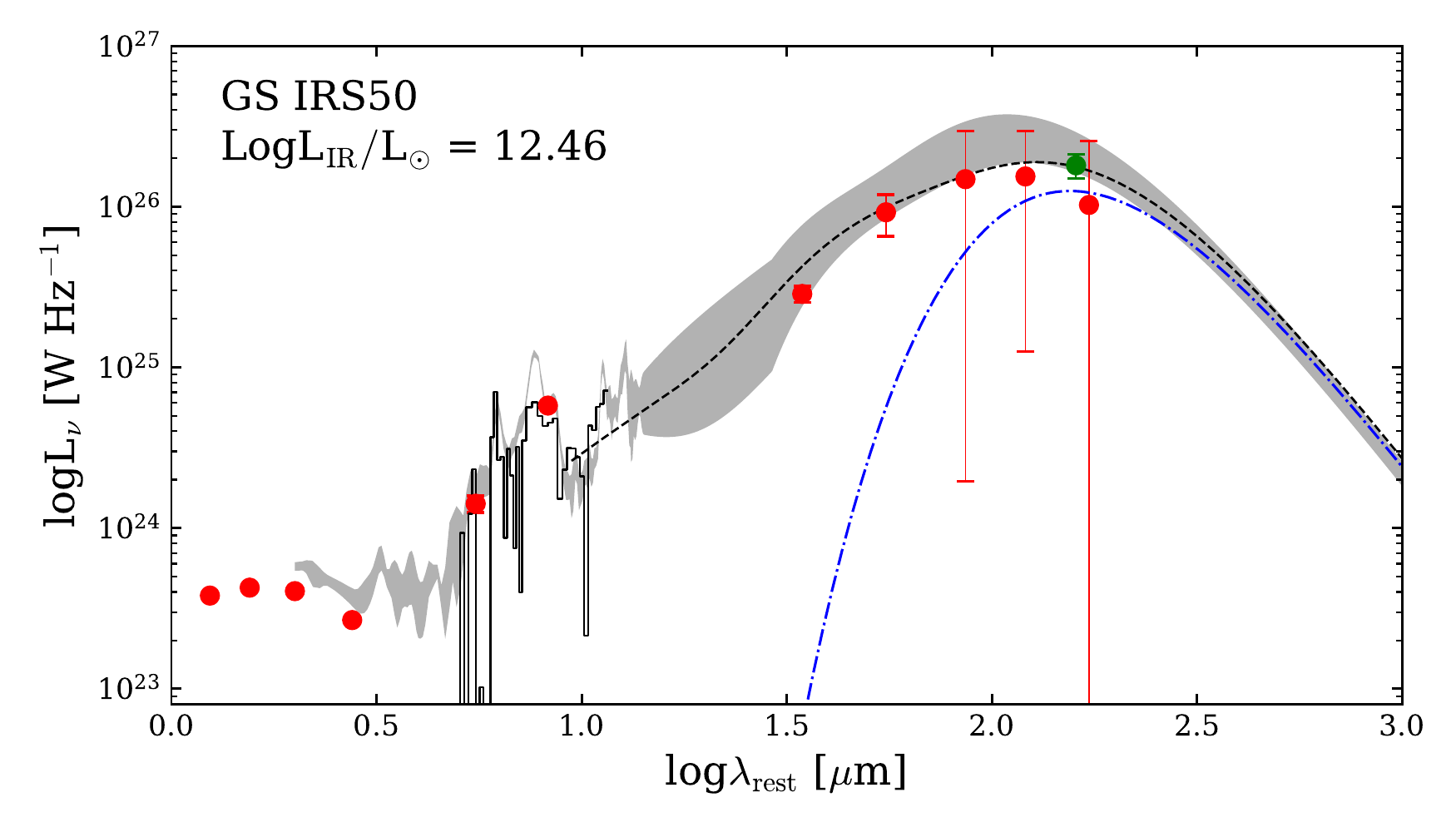}{0.49\textwidth}{}
          \fig{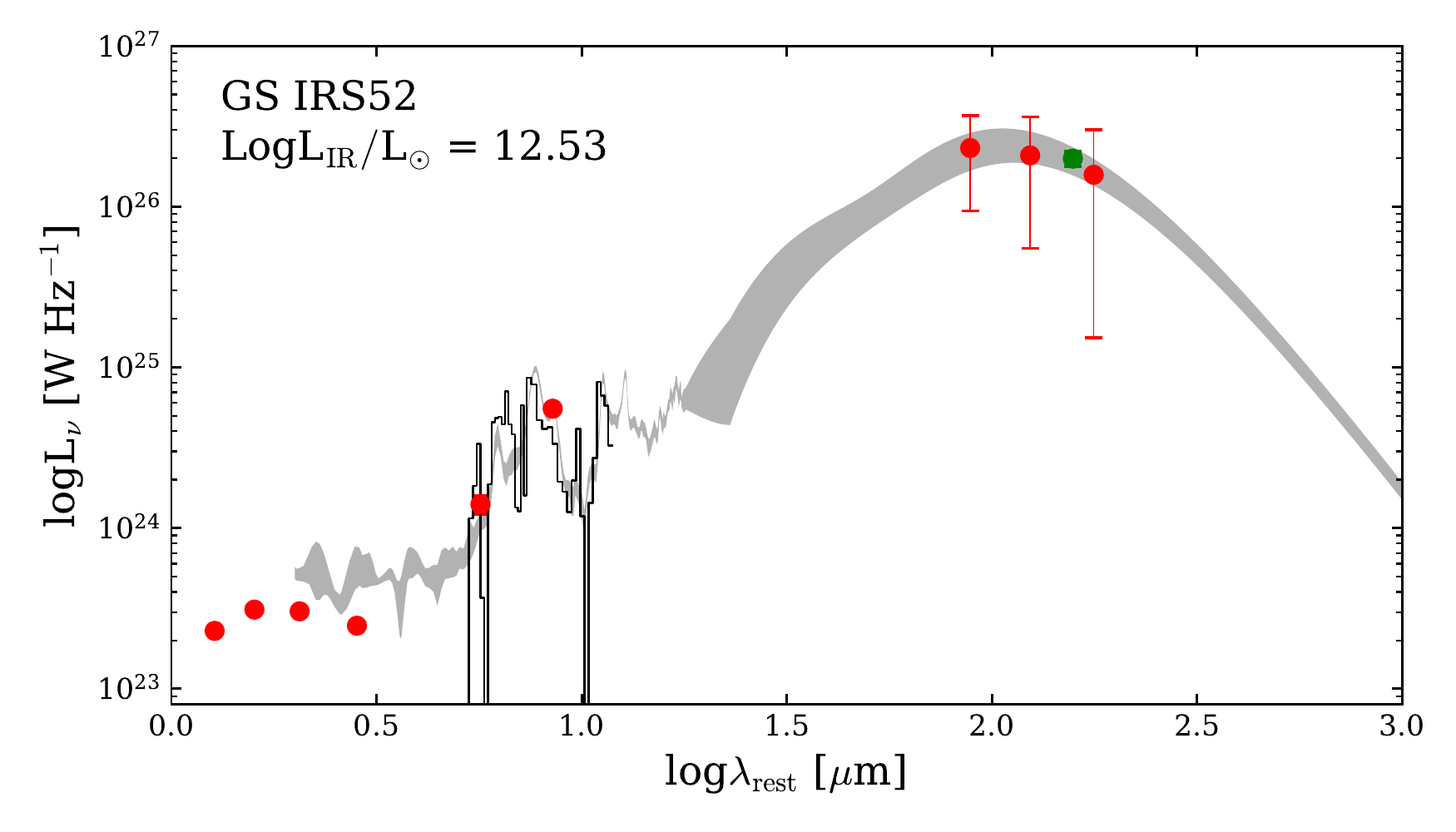}{0.49\textwidth}{}}
    \vspace{-2em}
   \gridline{\fig{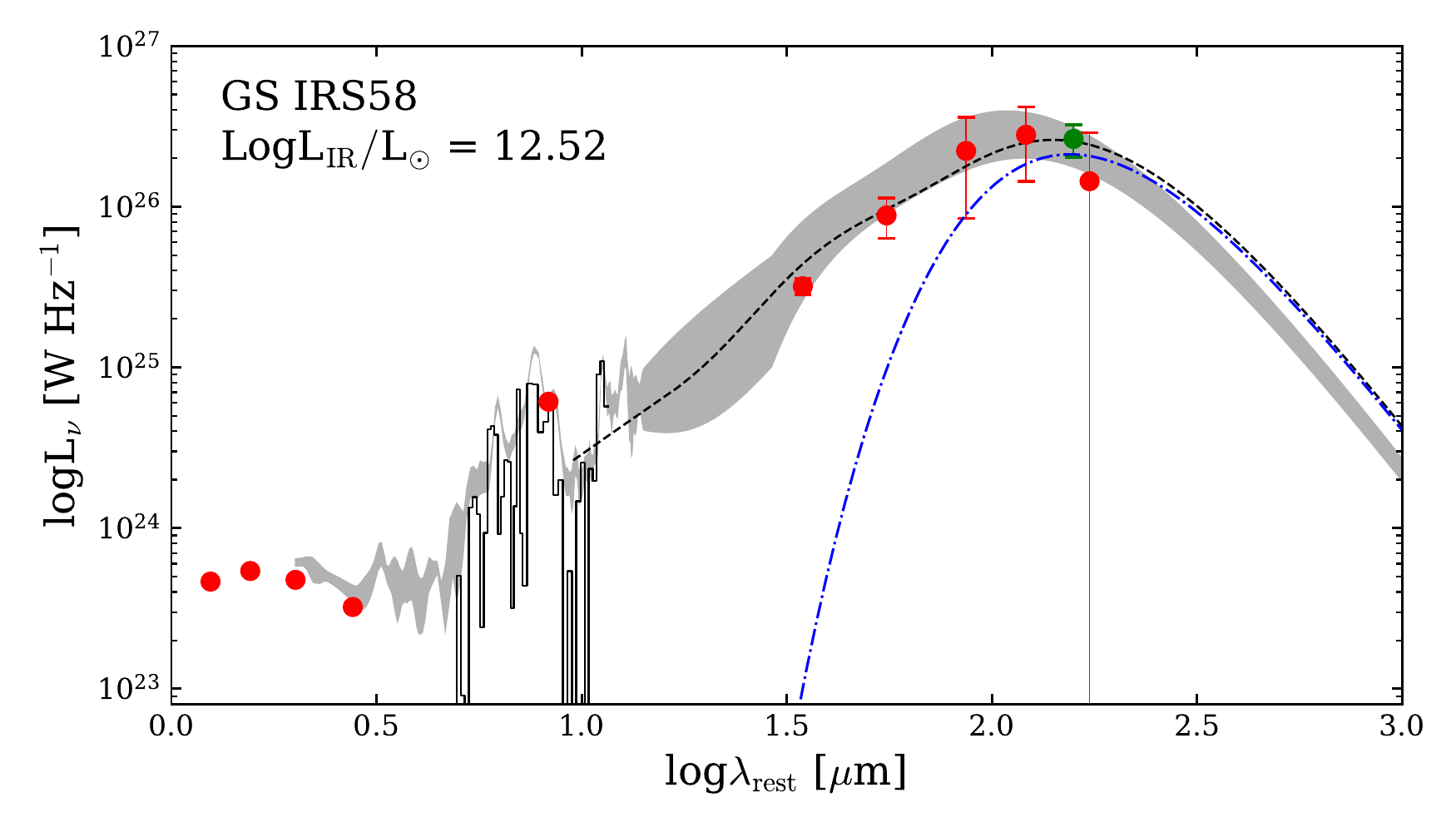}{0.49\textwidth}{}
          \fig{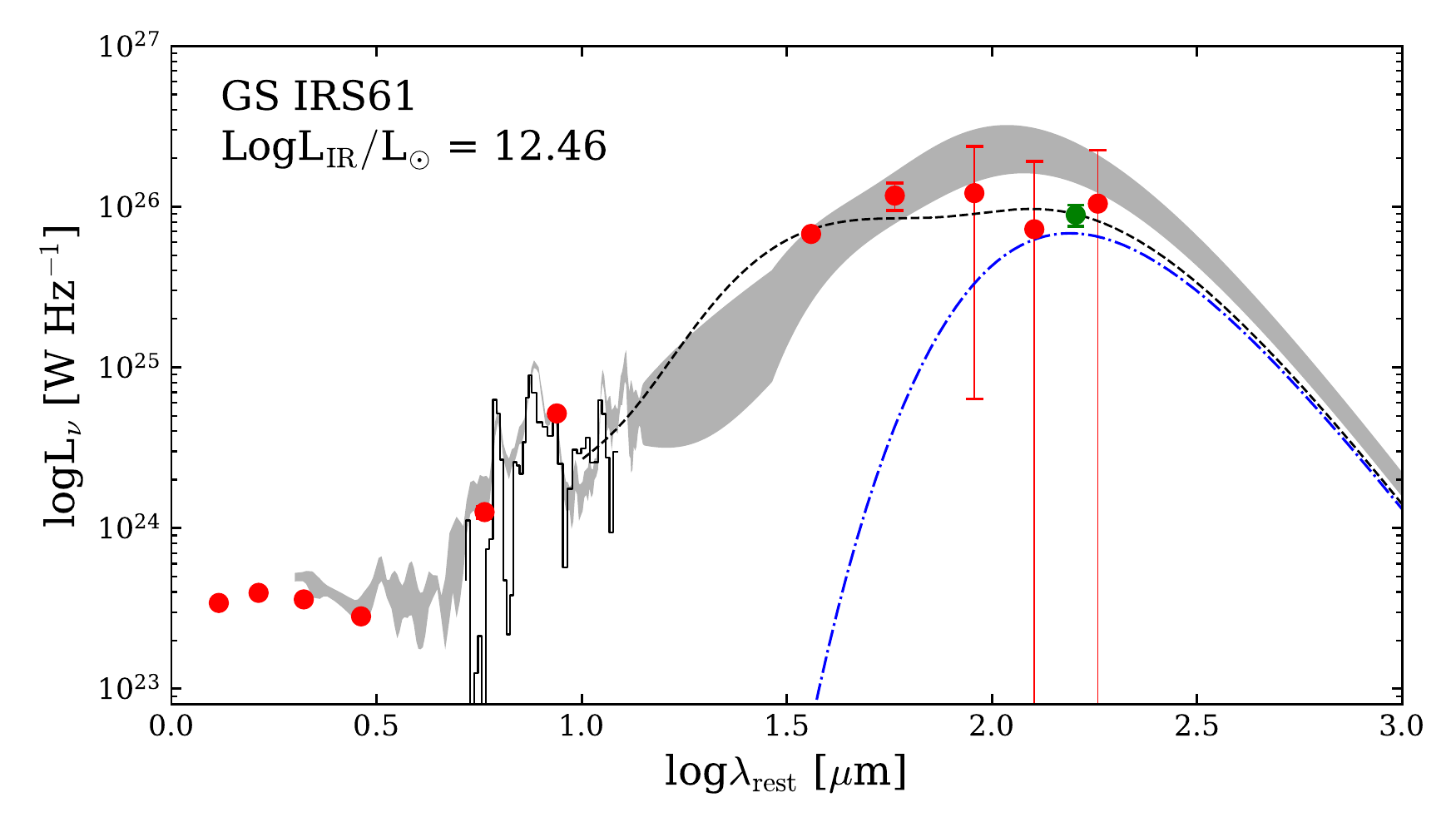}{0.49\textwidth}{}}
    \vspace{-1em}

    \caption{Multiwavelength photometry and spectra for each ALMA target. Shaded gray SEDs correspond to the best-fit $z\sim2$ template SED from the empirical library presented in \cite{Kirkpatrick2015}. The solid black line corresponds to a two-temperature modified blackbody + near-IR power-law fit, which we integrate to estimate \lir\ for each target. The dot-dashed blue line is the cold dust component from this fit which we integrate to calculate \lcold. Photometry in red correspond to \textit{Spitzer} and \textit{Hershel} observations. ALMA dust continuum is shown in green, and \textit{Spitzer} IRS spectra are over-plotted in black. We do not fit the blackbody + power-law models to GS IRS46 and GS IRS52 because these galaxies lack observations between rest-frame $30-70\,\mu$m. \label{seds}}
\end{figure*}


\subsection{Comparison Samples \label{comparisonSample}}

Since we have selected our $z\sim2$ sample to include only star-formation dominated systems, we emphasize literature comparison samples with comparable selections (EW$_{6.2\mu m}>0.5\,\mu$m, \citealt{Stierwalt2014}). For comparison with local (U)LIRGs, we use mid- and far-IR spectral line measurements from \cite{DiazSantos2013,DiazSantos2014,DiazSantos2017} and \cite{Stierwalt2014} for galaxies in the Great Observatories All Sky LIRG Survey (GOALS; \citealt{Armus2009}). To contextualize PAH and \cii\ line luminosities at lower \lir, we also compare our data to the intermediate$-z$ 5 mJy Unbiased Spitzer Extragalactic Survey (5MUSES; \citealt{Wu2010}), nearby galaxies from \cite{Sargsyan2014}, \cite{Magdis2014}, and \cite{Ibar2015}. To characterize the landscape of \cii\ observations at $z\sim2$, we also compare our \cii\ measurements to $z\sim2-3$ galaxies with  data from ALMA, APEX, or \textit{Herschel} FTS \citep{Ivison2010,Valtchanov2011,Schaerer2015,Gullberg2015,Zanella2018,Hashimoto2018,Rybak2019}. Prior observations of both PAH and [C II] in the same galaxy at $z\sim2$ are limited to a handful of systems observed with \textit{Spitzer} and the Redshift ($z$) and Early Universe Spectrometer (ZEUS) on the Caltech Submillimeter Observatory (CSO) \citep{Stacey2010, Brisbin2015}. 

For GOALS, 5MUSES and the ZEUS/CSO \cii\ sample, 6.2$\mu$m luminosities were derived using PAHFIT \citep{Smith2007} or CAFE \citep{Marshall2007}. It has been shown that PAHFIT-derived PAH line luminosities are greater than the those produced via continuum fitting methods by a factor of $\sim1.6-1.9$ for \lsix\ and \lii\ (e.g., \citealt{Sajina2007,Smith2007,Pope2008}). This is because PAHFIT is able to measure line emission in extended Lorentzian wings whereas continuum fitting methods do not. The $z\sim2$ \textit{Spitzer} IRS spectra do not have sufficient SNRs to use PAHFIT reliably, so we instead measure PAH lines using a continuum fitting technique described in Appendix Section \ref{pahmodel}. In summary, we fit a continuum + line model to isolated $6.2\,\mu$m and $11.3\,\mu$m regimes allowing the line strength, and galaxy redshift to vary. We also re-measure PAH luminosities in GOALS star-forming galaxies using our method, and divide PAHFIT values by a statistical conversion factor of 1.6 and 2.3 for \lsix\ and \lii\ respectively to match our quantities derived at higher redshift.

The GOALS sample is nearby and resolved by the \textit{Spitzer} IRS slit, which is centered on the nuclear region of each galaxy and will not capture the total mid-IR continuum and PAH flux \citep{Armus2009,Stierwalt2013}. For fair comparison with high-$z$ galaxies that are completely covered by the IRS slit, we correct the PAH line fluxes of GOALS using slit-corrections in \cite{Stierwalt2014} determined from the ratio of total \textit{Spitzer} IRAC 8 $\mu$m flux to total IRS 8 $\mu$m flux. These corrections have a median value of 1.14 and a negligible impact on the average value of GOALS galaxies in the diagnostic plots. 


\begin{figure}[t]
    \includegraphics[width=0.48\textwidth]{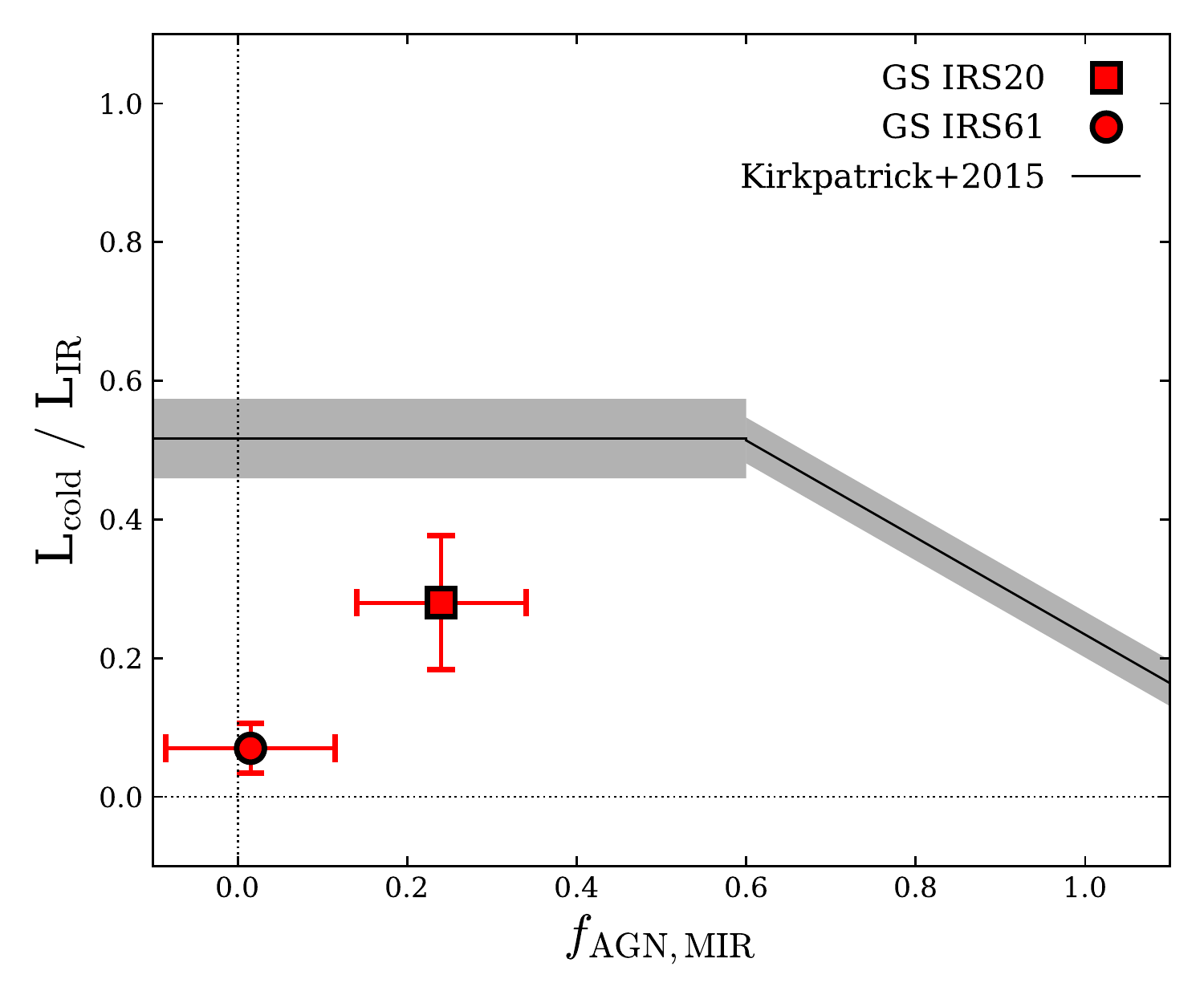}
    \caption{The ratio of cold dust emission, \lcold, to \lir\ as a function of \fagn, the AGN contribution to emission at mid-IR wavelengths. \lcold\ comes from integrating the cold dust component in our two-temperature SED fits shown in Figure \ref{seds}. Our data is shown in red. The solid black line indicates the best-fit trend for 343 (U)LIRGs between $z=0.3-2.8$ from \cite{Kirkpatrick2015}, with the $1\sigma$ uncertainty shaded in gray. \label{fig:fig_tmp}}
\end{figure}


\begin{figure}[t]
    \includegraphics[width=0.48\textwidth]{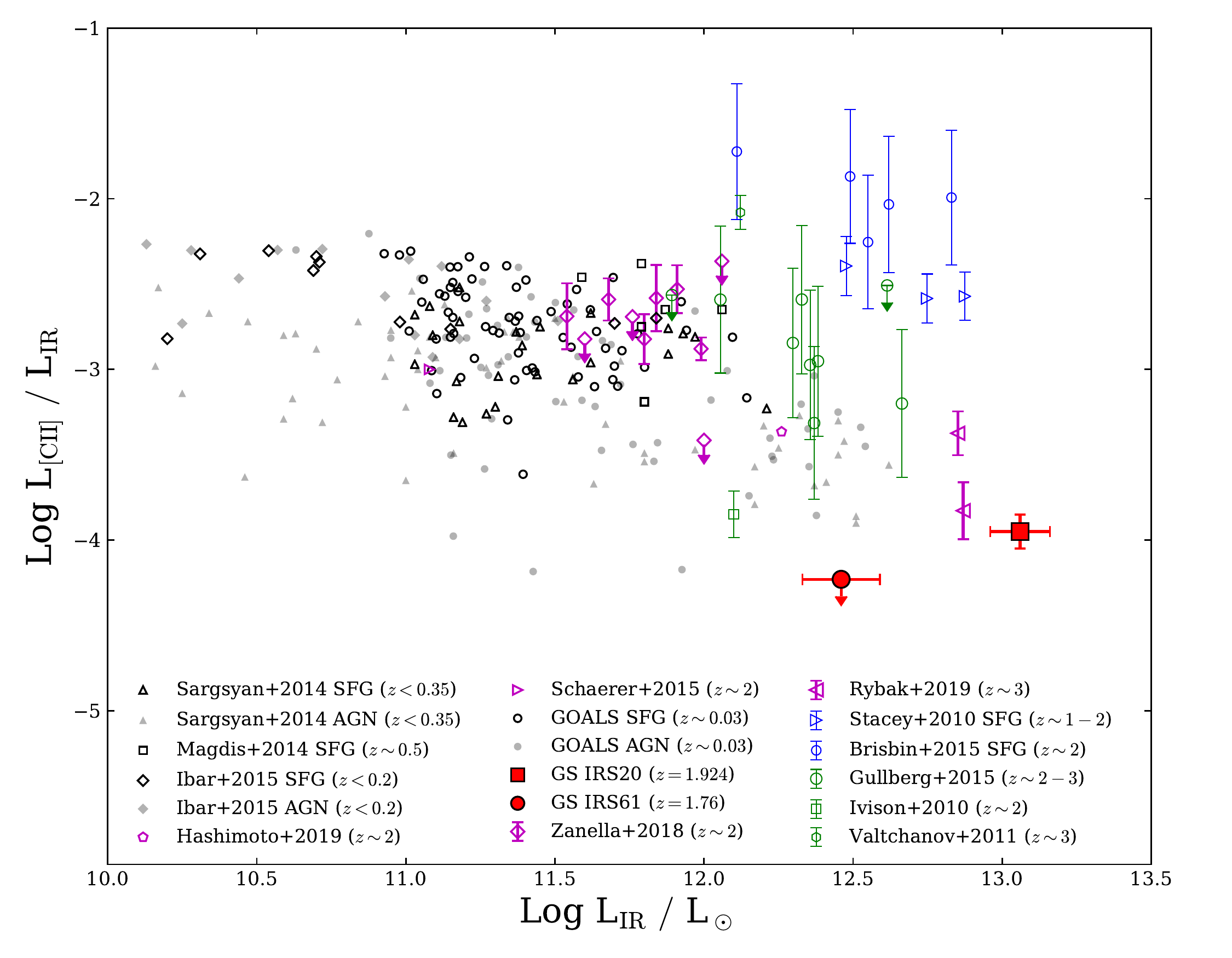}
    \caption{The ratio of \cii\ luminosity to \lir\ in low- and $z\sim2-3$ star-forming galaxies and AGN. We show local star-forming galaxies as black open symbols and low$-z$ AGN as gray symbols. At $z=2-3$, we show both star-forming and AGN systems with colored symbols: ALMA-derived \cii\ luminosities are shown in red (this work) and magenta. Blue symbols indicate \cii\ observations from ZEUS/CSO, which we re-calculate as described in Section \ref{ciiLineLums}. Green symbols correspond to galaxies targeted using APEX and \textit{Herschel}, including the lensed SPT DSFG sample of \cite{Gullberg2015}, which we de-magnify using their average magnification factor of 14.1. We include AGN in this figure to demonstrate the spread in \lcii/\lir\ observed in all galaxies; however, we emphasize that this work focuses on the range of star-formation properties in galaxies without AGN. \label{ciiLIR}}
\end{figure}


\subsection{SED Fits to near-IR through sub-mm Photometry}

Near-IR through sub-mm photometry are shown in Figure \ref{seds}. For comparison, we overplot the average SED of $z\sim2$ \loglir$\,=12.5$ star-forming galaxies from \cite{Kirkpatrick2015}, scaled to best match the observations. The excellent agreement at $5-15\mu$m is due to the fact that our galaxies are part of the sample used in generating the \cite{Kirkpatrick2015} templates, which were normalized in the mid-IR. 

To calculate total $8-1000\,\mu m$ IR luminosities, we fit a two-temperature modified blackbody + power-law model between the \textit{IRS} spectra at rest wavelengths above $9\ \mu$m out to the far-IR photometry, motivated by \cite{Kirkpatrick2015} who find that a two-temperature model yields good fits to the far-IR SEDs of $z=0.3-2.8$ (U)LIRGs. For all fits we keep the dust emissivity $\beta$ fixed to a value of $1.5$, and the temperature of the cold dust component fixed at T$_{cold}=26.1$ K corresponding to the average value of galaxies in the \cite{Kirkpatrick2015} sample with \fagn$\,\leq0.3$. From the fits, we measure \lir\ and the fraction of IR emission originating from the cold dust component (\lcold/\lir). Table \ref{sedfits} reports best-fit values for T$_{warm}$, the modified-Blackbody temperature of the warm dust component, and \lcold\ with their associated $1\sigma$ uncertainties for GS IRS20 and GS IRS61. GS IRS46 and GS IRS52 do not have rest-frame photometry between $30-70\,\mu$m, so we determine a best-fit template from the \cite{Kirkpatrick2015} library by matching to the available observations above rest-frame $9\,\mu$m. The scale-factors for each template are 11.7 and 2.6 in GS IRS46 and GS IRS52 respectively. We then integrate the scaled \cite{Kirkpatrick2015} template to calculate \lir\ in these two galaxies. 

Models fits are shown in Figure \ref{seds} as dashed black lines. \cite{Kirkpatrick2015} find that \lcold/\lir$\,\approx0.5$ on average for $z=0.3-2.8$ (U)LIRGs with $0<\,$\fagn$\,<0.6$. In GS IRS20, the one galaxy where \cii\ was detected at high significance, and GS IRS61, the target with a secure \cii\ upper limit, we measure \lcold/\lir$\,=0.28$ and $0.07$ respectively, at the extreme lower end of the distribution for galaxies of similar \fagn\ and \lir\ (Figure \ref{fig:fig_tmp}). GS IRS20 and GS IRS61 deviate from the mean of the \cite{Kirkpatrick2015} sample by $\approx2.5\sigma$ and $\approx12\sigma$ respectively.  Both systems have T$_{warm}$ comparable to stacked templates of similar \fagn\ (Table \ref{sedfits}), indicating that low \lcold/\lir\ is driven by an increase in the warm dust content of these two galaxies, and not a rise in the warm dust temperature. 


\begin{figure*}[t!]
     \gridline{\fig{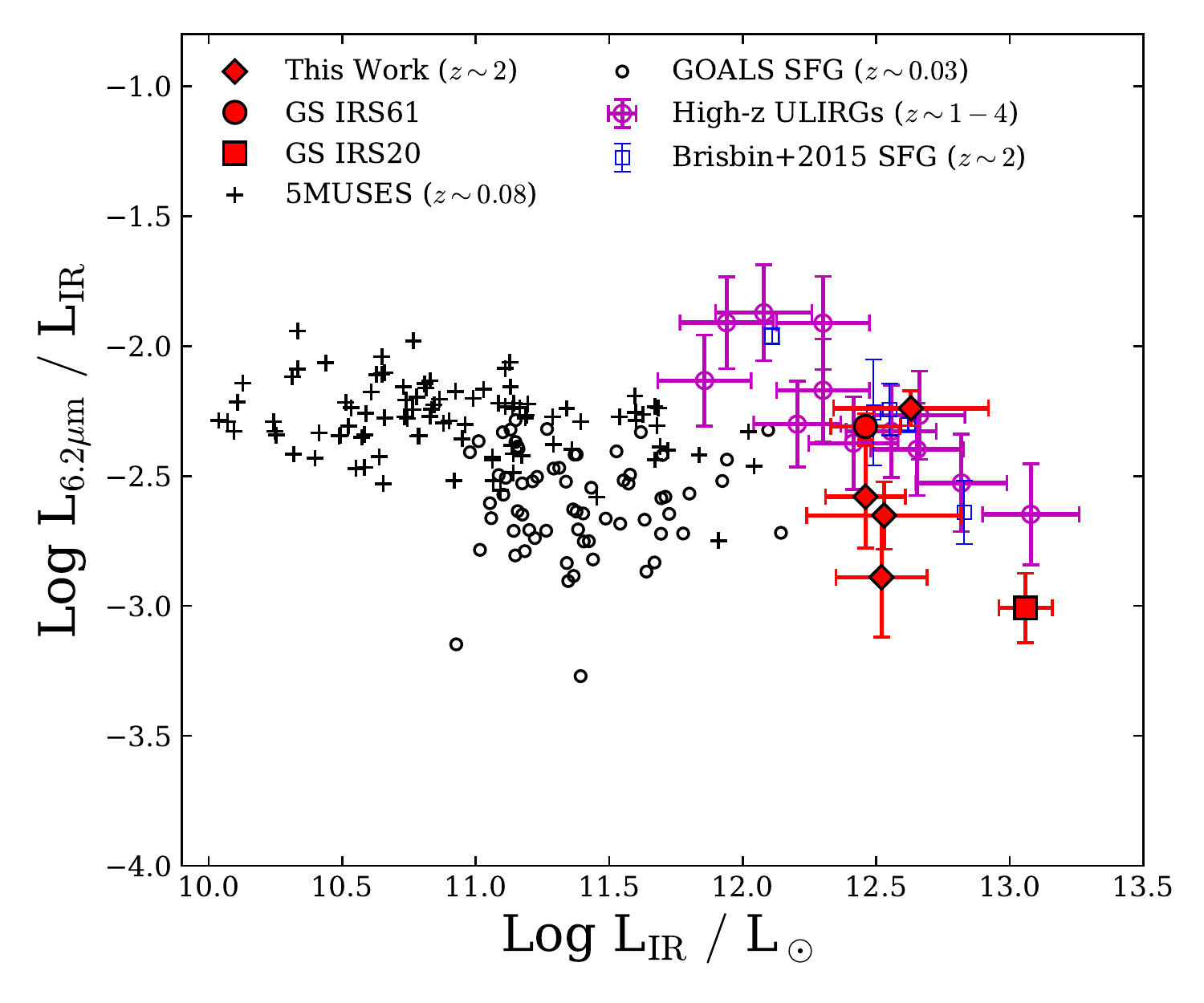}{0.49\textwidth}{}
          \fig{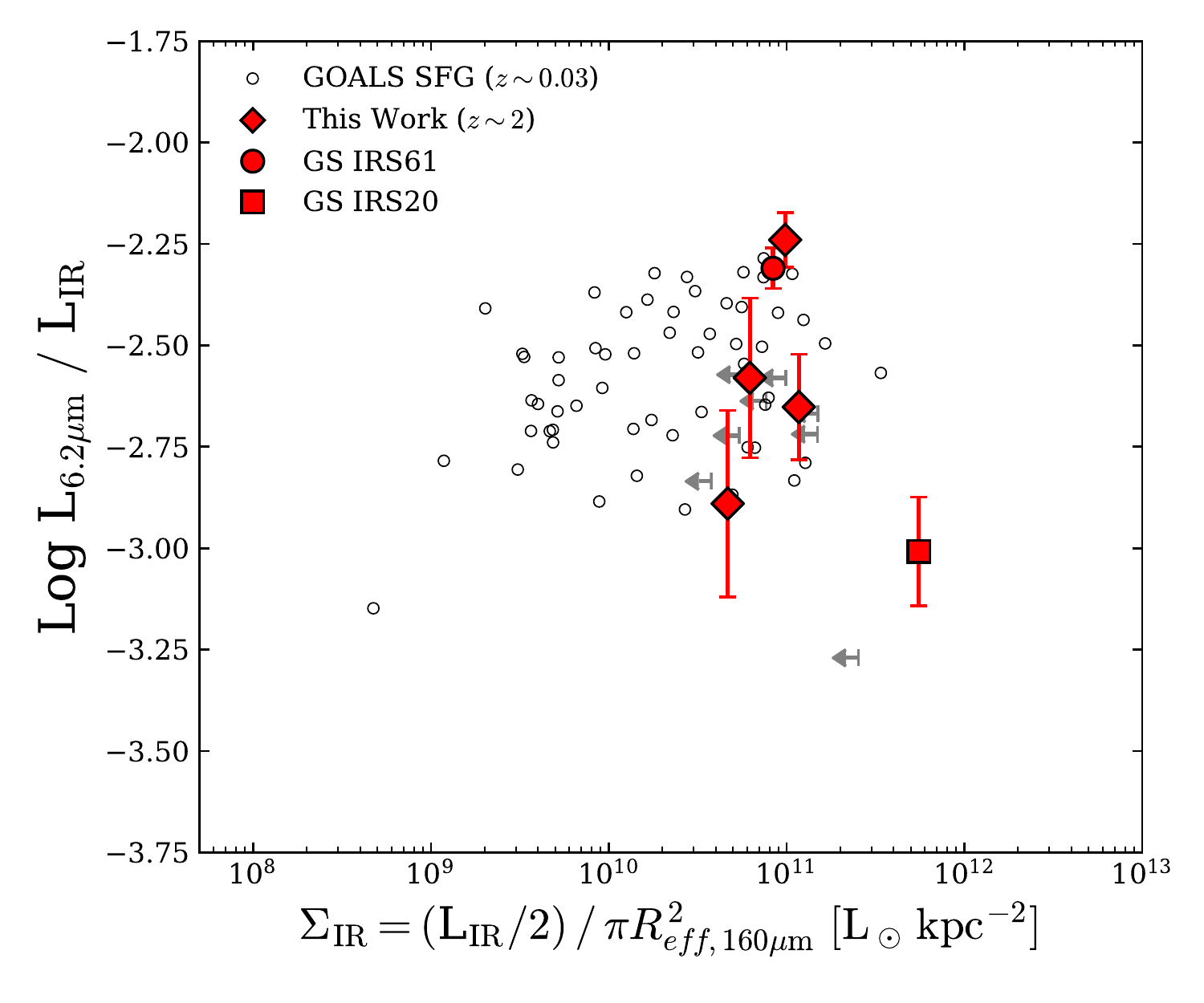}{0.49\textwidth}{}}
    \vspace{-20pt}
    \caption{(\textit{Left}) The ratio of \lsix\ to \lir\ in low- and high-redshift IR-luminous galaxies as a function of \lir. Local galaxies are taken from 5MUSES \citep{Wu2010} and the GOALS sample \citep{DiazSantos2013,DiazSantos2014,DiazSantos2017,Stierwalt2014} and follow the color scheme of previous figures. High redshift galaxies are represented with colored symbols: red corresponds to our sample, and blue indicates galaxies from \cite{Brisbin2015}. (U)LIRGs shown in purple are tabulated in \cite{Pope2013}. (\textit{Right}) The ratio of \lsix\ to \lir\ vs. IR surface density. Upper limits on \sigmaIR\ for GOALS galaxies smaller than the Herschel beam are shown in gray. The offset between low- and high-redshift galaxies observed in the \textit{Left} panel is removed when normalizing \lir\ by the cold dust surface density traced by rest-frame 160$\mu$m emission with ALMA and Herschel.   \label{lsixLIR}}
\end{figure*}


\section{Results}
\subsection{\cii\ Line Luminosities} \label{ciiLineLums}
Of six galaxies targeted with ALMA, we only detect the 158$\mu$m [\ion{C}{2}] fine-structure line in one galaxy, GS IRS20, the most IR-luminous source in our sample. For one other galaxy in our sample, GS IRS61, the \cii\ line was reliably covered by our ALMA observations ($p(l|\mathrm{ALMA},\Delta z)=93\%$). For the other targets, our ALMA observations have a probability $>40\%$ that we missed the redshifted \cii\ line given prior redshift uncertainties and the sparse frequency coverage of ALMA spectral windows in Band 9. For the remainder of the paper we only include GS IRS20 and GS IRS61 in any analysis that involves \cii.

Figure \ref{ciiLIR} shows the \cii\ deficit for low-redshift (U)LIRGs and star-forming galaxies at $z\sim2-3$ from this sample and the literature. We note that the number of IR-luminous galaxies at $z\gtrsim4$ with \cii\ detections is growing\footnote{E.g., \citealt{Gallerani2012,Walter2012,Riechers2013,Bussmann2013,Rawle2014,DeBreuck2014,Maiolino2015,Capak2015,Oteo2016,Pentericci2016,Carniani2017,Jones2017,Matthee2017,Smit2018,Carniani2018,Gullberg2018,Decarli2018,Hashimoto2018,LeFevre2019,Tadaki2019,Hashimoto2019}}; however, we restrict our current analysis of the high$-z$ landscape to $z\sim2-3$ to focus on galaxy properties near the cosmic star-formation rate density peak. The ratio of \lcii\ to \lir\ in GS IRS20 is comparable to other $z\sim3$ ALMA \cii\ detections from \cite{Rybak2019} and possibly consistent with the extrapolation of the low$-z$ \cii-deficit to \loglir$\,\geq12.5$. GS IRS61 is $\sim2$ dex below GOALS star-forming galaxies of \loglir$\,\approx12$. 

There is a significant offset on the order of $0.5-1.5$ dex between \lcii/\lir\ at $z=2-3$  found with ALMA and those reported by \cite{Stacey2010} and \cite{Brisbin2015} using ZEUS/CSO (blue in Fig. \ref{ciiLIR}). The spectral resolution of ZEUS is $150-300$ km s$^{-1}$, comparable to the expected line-width of \cii\ emission in some cases, making the flux measurements sensitive to the number of spectral pixels included when integrating a low SNR line. We re-calculate all ZEUS/CSO \cii\ luminosities using only the peak pixel flux assuming a line-width of $150-300$ km s$^{-1}$. After these corrections, the $0.5-1.5$ dex offset between ALMA and ZEUS observations in $z\sim2$ SFGs persists. There are multiple factors that could contribute to this offset, including physical variations in \cii/\lir\ with star-formation rate surface density (e.g., \citealt{Smith2017,DiazSantos2017}), or observational limitations such as large beam sizes, lower spectral resolution, and flux calibration uncertainties on the order of 30\% \citep{Brisbin2015}. 

\begin{deluxetable}{lccc}
    \tablecaption{Derived Parameters from IR-SED Fits\label{sedfits}}
    \tabletypesize{\footnotesize}
    \tablehead{Target & T$_{cold}$ [K] & T$_{warm}$ [K]&  L$_{cold}$/\lir }
    \startdata 
    GS IRS20 & $26.1$ (fixed)\tablenotemark{a}  & $57\pm1$  &  $0.28\pm0.10$ \\[.3ex]
    GS IRS61 & $26.1$ (fixed)\tablenotemark{a} & $59\pm2 $ &  $0.07\pm0.04$  \\[.3ex]
    \hline \\[-13pt]
    MIR0.0\tablenotemark{b} & $25.7\pm0.6$ & $66\pm2$  & $0.51\pm0.04$ \\[.3ex]
    MIR0.2\tablenotemark{b} & $24.6\pm1.3$ & $62\pm1$  & $0.44\pm0.06$ \\[.3ex]
    \enddata
    \tablenotetext{a}{T$_{cold}$ was fixed in the SED fits of both GS IRS20 and GS IRS61 to the average of galaxies in the \cite{Kirkpatrick2015} sample with \fagn$\leq0.3$}
    \tablenotetext{b}{Empirical templates from \cite{Kirkpatrick2015} of comparable \fagn\ to GS IRS20 and GS IRS61. MIR0.0 and MIR0.2 correspond to \fagn$\,=0.0$ and \fagn$\,=0.2$ respectively.}
\end{deluxetable}


\subsection{PAH Properties}

The relationship between PAH emission and dust emission evolves with redshift and is likely related to a number of factors, including \fagn, SFR, and the number of PDRs per unit molecular gas mass \citep{Smith2007,Pope2013}. Star-forming galaxies in our sample at $z\sim2$ have 6.2 $\mu$m PAH luminosities $0.3$ dex brighter than local (U)LIRGs of comparable \lir\ after accounting for the differences in \lsix\ measurement techniques (see Section \ref{comparisonSample}), but follow a deficit in PAH emission towards higher \lir\ whose magnitude of decline is equal to or greater than the deficit between other far-IR fine-structure lines and \lir\ \citep{Pope2008,Pope2013,Sajina2008,GC2011,Stierwalt2014,Shipley2016,Cortzen2019}. In addition to being a function of \lir, \lsix/\lir\ also changes with $z$ (e.g. \citealt{Pope2013}), as demonstrated in Figure \ref{lsixLIR} (\textit{Left}) which shows the ratio of \lsix\ to \lir\ for low- and high-$z$ star-forming galaxies. Galaxies at $z\sim2$ in our sample are brighter in \lir\ by a factor of $\gtrsim0.5$ dex compared to low-$z$ (U)LIRGs of comparable \lsix/\lir; changes in either/both of \lsix\ and \lir\ could drive the difference between low-redshift and $z\sim2$ galaxies in Fig. \ref{lsixLIR}. In any case, this trend persists if we instead use the ratio of 11.3 $\mu$m PAH luminosity to \lir\ as well as values for \lsix\ in GOALS measured using our method described in Section \ref{pahmodel}. 

\cite{DiazSantos2017} show that the IR surface density is a good predictor of physical PDR conditions such as gas density and incident radiation field strength. Furthermore, spatially resolved studies of nearby and $z\sim3$ star-forming galaxies have shown the star-formation rate surface density (\sigmaSFR) to be a major driver of the \cii-deficit \citep{DiazSantos2014,Smith2017,Rybak2019}. In light of these results, and without spatial information at shorter wavelengths more aptly suited for tracing \sigmaSFR, we calculate \sigmaIR$\,=($\lir$/2)/\,\pi R^2_{\mathrm{\footnotesize eff},160}$, the effective IR surface density using \Reff\ as measured with ALMA for our sample and \textit{Herschel} PACS in GOALS. Figure \ref{lsixLIR} (\textit{Right}) demonstrates that the offset between high- and low-$z$ galaxies in \lsix/\lir\ disappears when plotted against \sigmaIR. 


\begin{figure}[t!]
    \includegraphics[width=0.48\textwidth]{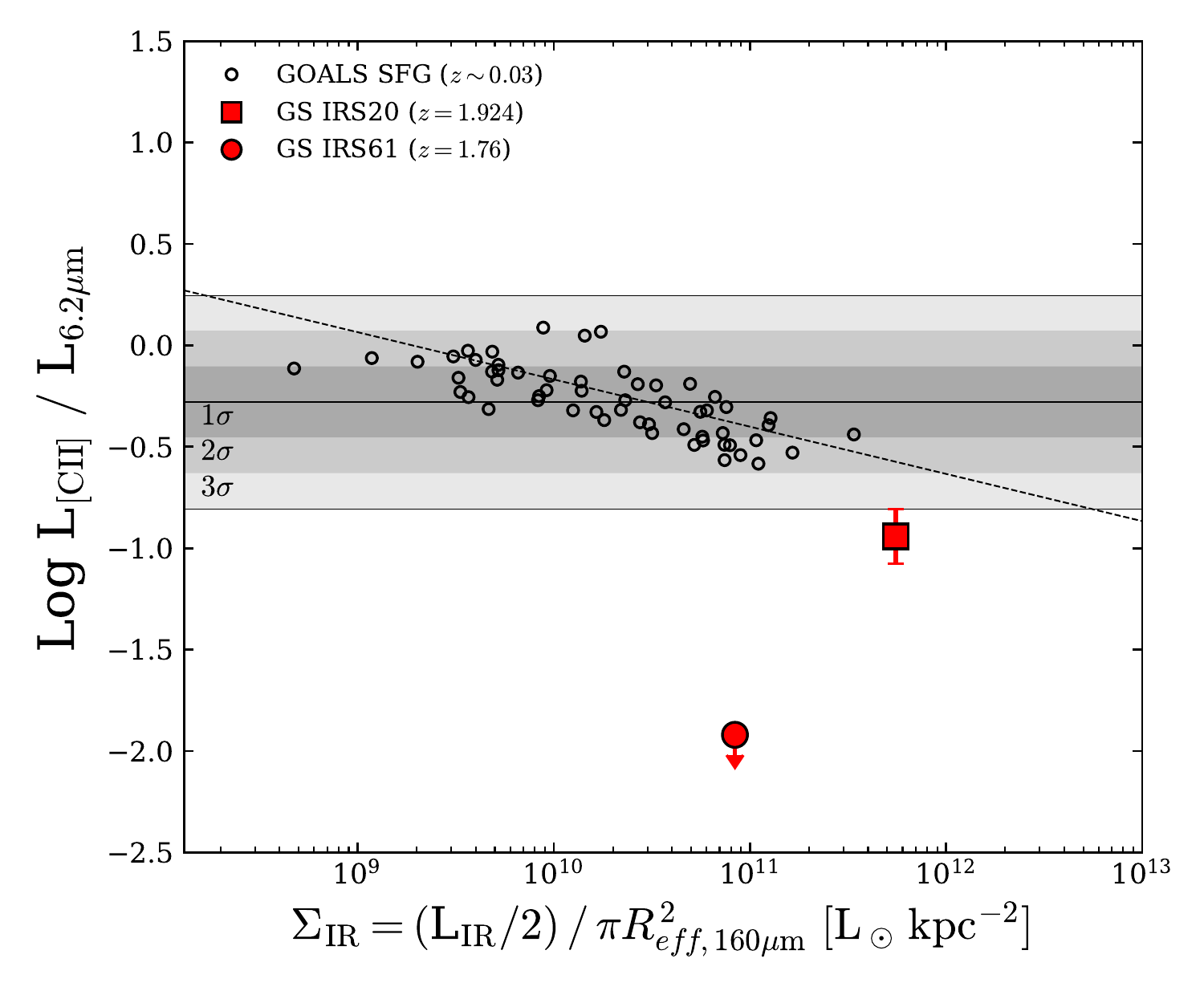}
    \caption{The ratio of \cii\ luminosity to 6.2$\mu$m PAH luminosity in low-redshift and $z\sim2$ IR-luminous star-forming galaxies as a function of IR surface density. The effective radius for all sources shown is calculated at rest-frame $160\,\mu$m continuum. The gray shaded regions contain the $1\sigma$, $2\sigma$, and $3\sigma$ dispersions around the mean of \lcii/\lsix\ in star-forming GOALS galaxies. The dotted black line corresponds to the best-fit trend in GOALS, and has a slope of $-0.23\pm0.08$ and zero-point equal to $2.2\pm0.9$. The dearth of high$-z$ points on this Figure demonstrates the need for more observations of \cii, PAH, and IR size measurements in the same galaxies. \label{ciiL62}}
\end{figure}


\begin{figure}
    \includegraphics[width=0.48\textwidth]{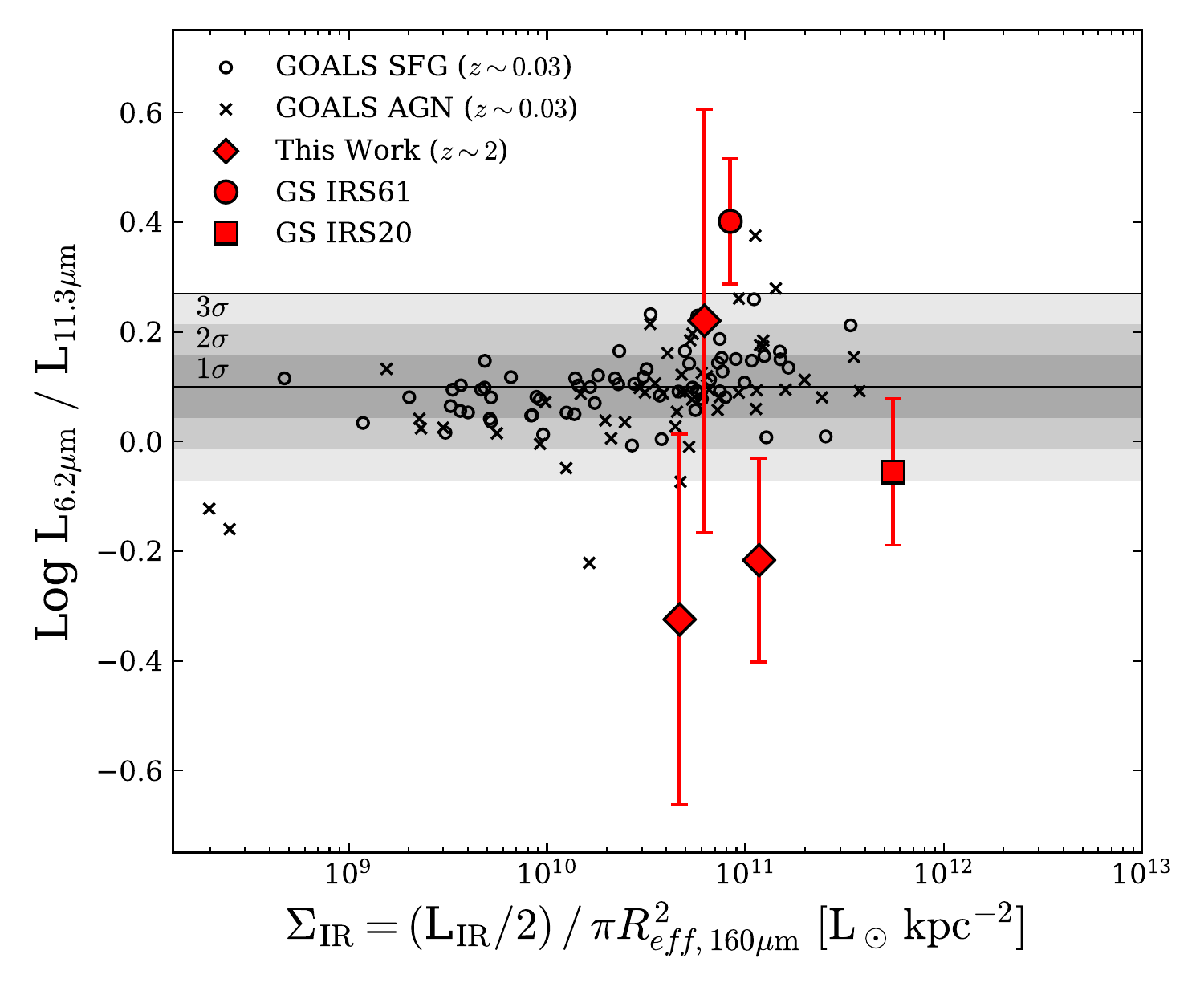}
    \caption{The ratio of \lsix\ to \lii\ vs. effective IR surface density calculated at rest-frame 160 $\mu$m. The color scheme follows previous figures. \lsix/\lii\ does not change drastically with \sigmaIR\ in star-forming GOALS galaxies, with minimal $1\sigma$ scatter $\sim0.06$ dex shaded in gray about an average value of $0.08$. \label{pahSigma}}
\end{figure}


\subsection{The Ratio of \cii\ to PAH Luminosity}

Figure \ref{ciiL62} shows \lcii/\lsix\ as a function of \sigmaIR\ for our sample, and local (U)LIRGs. GOALS star-forming galaxies (EW$_{6.2\mu m}\geq0.5\,\mu$m) have a tight ratio of \lcii/\lsix\ with a dispersion of 0.18 dex, less than the $\sim0.3$ dex dispersion observed in both \lcii/\lir\ and \lsix/\lir. We fit a linear relation to star-forming GOALS galaxies on Figure \ref{ciiL62}, and find that \lcii/\lsix\ and \sigmaIR\ anti-correlate with a slope of $-0.23\pm0.08$ and zero-point $2.2\pm0.9$. Although high-redshift observations remain limited by small sample statistics, the spread in \lcii/\lsix\ between GS IRS20 and GS IRS61 at $z\sim2$ is 0.98 dex, five times greater than the  dispersion of local star-forming (U)LIRGs, an observation that holds regardless of how the PAH luminosities are measured in GOALS (see Section \ref{comparisonSample}).

GS IRS20 and GS IRS61 are $\sim0.67$ and $\geq1.65$ dex respectively below the mean of \lcii/\lsix\ observed in GOALS star-forming galaxies, after accounting for the differences in how the PAHs were measured. GS IRS20 is possibly consistent with the extrapolation of the low$-z$ negative trend between \lcii/\lsix\ and \sigmaIR\ beyond the most compact GOALS star-forming galaxy; however, this cannot explain the extremely low ratio observed in GS IRS61. GS IRS61 shows no indication of a deeply buried AGN (Figure \ref{irsSpectra}), and there is a low probability that we missed the redshifted \cii\ line (see Section \ref{linesearch}). The dust continuum is marginally more extended than the ALMA beam and would have to be extraordinarily compact (\sigmaIR\ would have to increase by $>2$ orders of magnitude) to be consistent with the extrapolated low$-z$ trend. For these reasons, GS IRS61 is likely a highly unusual source when compared to low$-z$ star-forming galaxies of comparable \sigmaIR.

If we assume that the relevant physical parameters of the $z\sim2$ galaxies are drawn from the same distribution that is observed in GOALS, then GS IRS20 and GS IRS61 would be $\sim3\sigma$ and $\gtrsim6\sigma$ below the low$-z$ mean. The likelihood of observing two galaxies at $3\sigma$ and $6\sigma$ from the norm is $\approx10^{-11}$. Therefore, the offset in \lcii/\lsix\ between low-$z$ (U)LIRGs and what we measure in our sample may relate to changes in the physical ISM conditions. 

\section{Discussion}
\subsection{PAH Heating vs. Far-IR Cooling}

We find a difference in the ratio of \cii\ to PAH emission between local ULIRGs and observations of two $z\sim2$ dusty star-forming galaxies including one upper limit (Fig. \ref{ciiL62}), which could be due to changes in heating and cooling mechanisms. As opposed to being scaled up versions of nearby star-forming galaxies, starbursts at earlier times may exhibit evolution in their ISM conditions. While the behavior of both PAH and \cii\ changes at a low metallicity \citep{Shivaei2017,Croxall2017}, we do not expect this to affect our massive ($\log\,$M$_*/$M$_\odot\sim11$) $z\sim2$ galaxies given the $z\sim2$ mass-metallicity relation \citep{Sanders2015}. 

PAHs and other small grains are important sources of photoelectrons in PDRs (e.g., \citealt{Bakes1994}), and the ratio of far-IR line to PAH emission is sensitive to the photoelectric heating efficiency (\ephot) of the PDR gas. As noted by \cite{Helou2001}, \lpah\ (or \lsix) may be more appropriate normalization factors for \lcii\ than \lir\ given the direct relationship with \ephot:
\begin{equation}
\frac{\mathrm{L}_\mathrm{\tiny[C\ II]}+\mathrm{L}_\mathrm{\tiny[O\ I]}}{\mathrm{L}_\mathrm{\tiny PAH}} = (\eta_{\mathrm{\tiny[C\ II]}} + \eta_{\mathrm{\tiny[O\ I]}}) \epsilon_{\mbox{\tiny PE}}\approx \frac{\Gamma_{\tiny\mbox{gas}}^{e^{-}}}{\Gamma_{\tiny\mbox{dust}}^{\tiny\mbox{PAHs}}}
\label{ephot}
\end{equation}
where following \cite{Croxall2012}, $\eta_{\mbox{\tiny[C\ II]}}$ and $\eta_{\mbox{\tiny[O\ I]}}$ represent the relative contribution of the two principal cooling channels to the total gas cooling. $\Gamma_{\tiny\mbox{gas}}^{e^{-}}$ is the total gas heating via photoelectrons, and $\Gamma_{\tiny\mbox{dust}}^{\tiny\mbox{PAHs}}$ the total dust heating accounted for by PAHs. Cooling from other far-IR such as [\ion{C}{1}] and [\ion{Si}{2}] are assumed to be negligible (i.e., $\eta_{\mbox{\tiny[C\ II]}} + \eta_{\mbox{\tiny[O\ I]}}\sim1$). 

Assuming that the 6.2$\mu$m PAH feature linearly scales with total PAH luminosity (e.g., \citealt{Smith2007}), and the fraction of \cii\ emission originating from PDRs is roughly constant, then the ratio \lcii/\lsix\ probes the difference of photoelectric efficiency and normalized cooling via [\ion{O}{1}]. Knowing that the ratio of [\ion{O}{1}] emission to \cii\ emission of PDR origin varies by an order of magnitude in nearby (U)LIRGs \citep{DiazSantos2017}, the location of GS IRS20 and GS IRS61 on Fig. \ref{ciiL62} could be interpreted as evidence for enhanced \oi\ cooling in these galaxies if the total \ephot\ is constant. \cite{DiazSantos2017} demonstrate that \oi/\cii\ correlates with gas and dust temperature within PDRs, and \oi/\cii$\,>1$ where dust temperatures exceed $\sim35$ K. Indeed, warm-dust blackbodies (T$_{warm}\sim60$ K) dominate the IR SEDs of both GS IRS20 and GS IRS61 (\lcold/\lir$\,\lesssim0.3$, Table \ref{sedfits}), consistent with enhanced PDR cooling through \oi\ emission. Moreover, our $z\sim2$ sample has high \sigmaIR\ compared to the average of GOALS (Fig. \ref{lsixLIR} \textit{Right}), implying more star-formation in smaller volumes. In such physical conditions, PDR densities are expected to be higher and exposed to more intense radiation fields where \oi\ naturally arises as the dominant cooling channel \citep{DiazSantos2017}. If the positions of GS IRS20 and GS IRS61 on Fig. \ref{ciiL62} are solely due to enhanced \oi\ cooling (\ephot$\,=\,$constant), then we calculate \loi$\,=7\times10^9$ L$_\odot$ and \loi$\,\gtrsim10^{10}$ L$_\odot$ for these two galaxies respectively in order to bring both in line with the GOALS sample. In this scenario, \loi/\lcii$\,\sim5$ in GS IRS20 and \oi\ dominates far-IR line cooling in both galaxies. 

Alternatively, low \lcii/\lsix\ could indicate a low \ephot\ by Equation \ref{ephot} if \oi\ emission is not significantly enhanced in GS IRS20 and/or GS IRS61. We speculate that a decrease in the photoelectric efficiency in high-$z$ dusty star-forming galaxies could play a role in enhancing star-formation rates compared to the galaxy main-sequence by reducing the coupling efficiency between interstellar radiation fields and gas heating. In other words, the colder ISM phases become less susceptible to temperature increases via stellar feedback as the reservoir of electrons in PAHs is diminished. Consequently, galaxies above the main-sequence would not exhibit strong far-IR line cooling at higher star-formation rates, as has been observed locally and tentatively at high$-z$ \citep{DiazSantos2017,Zanella2018}. A comprehensive study of far-IR fine-structure emission lines combined with mid-IR PAH spectra is needed to test this hypothesis, and the nature of gas heating and cooling at $z\sim2$ will be a function of \ephot, $\eta_{\mbox{\tiny[C\ II]}}$, and $\eta_{\mbox{\tiny[O\ I]}}$. Systematically low \ephot\ in dusty star-forming galaxies at $z\sim2$ would be associated with \oi/\cii$\,\sim1$ in a statistical sample controlled for \fagn, whereas \oi/\cii$\,>1$ would favor higher density PDRs with more \oi\ cooling. These far-IR cooling line ratios will be key for accessing the physical conditions in which most of the Universe's stellar mass was formed.


\subsection{Differences between GS IRS20 and GS IRS61}

The data in hand portrays an interesting dichotomy of ISM conditions between GS IRS20 and GS IRS61. A $1$ dex difference in \lcii/\lsix\ exists between the two galaxies, and is likely a function of PAH ionization state and therefore \ephot. Whereas the 6.2 $\mu$m feature traces ionized PAHs, the 11.3 $\mu$m complex arises from neutral PAHs yet to lose their surface electrons \citep{Tielens2008}. As a result, the ratio of \lsix/\lii\ is sensitive to the PAH ionization fraction in a galaxy, and also changes in the grain size distribution as observed near the nuclei of AGN \citep{Smith2007,Tielens2008}. Figure \ref{pahSigma} shows this ratio as a function of \sigmaIR\ for GOALS and our sample at $z\sim2$. GS IRS20 has a PAH line ratio near the local average, as may be expected if star-formation in this merging galaxy is proceeding in a comparable manner to what is found in GOALS, which are mostly mergers themselves. On the other hand, GS IRS61 has the highest ratio of \lsix/\lii\ amongst galaxies at low- and high-redshift. This is consistent with the location of GS IRS61 in Fig. \ref{ciiL62}: an increase in PAH ionization would lower \ephot, decoupling PAH and \cii\ emission to produce the extreme deficit in \lcii/\lsix\ observed. 

The only low-$z$ galaxies within $1\sigma$ of GS IRS61 on Fig. \ref{pahSigma} are a handful of GOALS AGN, and the ratio of \lsix/\lii\ appears larger than most star-forming GOALS galaxies, even after correcting for PAH extinction (see Fig.~2 of \citealt{Stierwalt2014}). Whether or not this is common at high-redshift remains to be explored; however, the scatter in \lsix/\lii\ we measure at $z\sim2$ is nearly three times larger than what is seen in the GOALS star-forming sample, although we note that error bars at higher $z$ are large. While both GS IRS20 and GS IRS61 have comparable far-IR colors, GS IRS61 has a lower \lcold/\lir\ (Fig. \ref{fig:fig_tmp}), indicating warmer dust conditions dominating the galaxy, consistent with low \ephot\ as larger dust grains absorb more of the incident radiation field in PDRs. The parameter space of PAH line ratios at cosmic noon has yet to be statistically explored, and may prove key for our understanding of dust properties and the link between stellar radiation fields and the ISM at the peak epoch of galaxy evolution.

\subsection{Future Outlook}

Testing the nature of gas heating and cooling in the ISM of high-redshift galaxies will be possible with future ALMA observations targeting \cii\ in IRS sources. Mid-IR spectra are crucial for constraining \fagn, from which the properties of star-formation at high$-z$ can be reliably characterized in the absence of or presence of an AGN. \textit{Spitzer's} cryogenic lifetime has ended, so the number of galaxies with available mid-IR spectra is currently limited. Future surveys with \textit{JWST}/MIRI will re-open the mid-IR Universe at high spectral sensitivity. 

Pending the launch of \textit{JWST}, ALMA can continue targeting IRS galaxies to explore the relationship between \cii\ and PAH emission as a function of \lir\ and \fagn. Understanding the intrinsic scatter in these relations will be crucial when designing efficient surveys that maximize the science potential of \textit{JWST}, and key for understanding the physics of gas heating and cooling in the early Universe, which observations with future facilities like \textit{Origins Space Telescope}\footnote{\url{https://origins.ipac.caltech.edu/}} or \textit{SPICA}\footnote{\url{https://spica-mission.org/}} will revolutionize. 


\section{Summary and Conclusions}

We have observed \cii\ emission in a sample of $z\sim2$ star-forming galaxies with existing detections of PAH dust emission in order to explain the balance of heating and cooling in the ISM and how it may be different from $z\sim0$. Our main conclusions are as follows : 

\begin{enumerate}
    \item We detect the dust continuum near the peak of the IR SED ($\lambda_{\tiny\mathrm{rest}}\sim160\ \mu$m) in all six targets. After correcting for known astrometry offsets between ALMA and \textit{HST}, the position of the dust continuum emission coincides with the rest-frame optical light in all but GS IRS46. Our most luminous target GS IRS20 is classified as a merger and is a clear starburst on the main-sequence diagnostic diagram. 
    \item We detect \cii\ in one target, GS IRS 20 at high SNR of 34.  The bright \cii\ emission and interesting optical morphology makes this an excellent target for follow-up ALMA observations to study its gas dynamics at higher spatial resolution. We place a deep upper limit on \lcii\ in one other galaxy, GS IRS61, after  calculating the probability the redshifted \cii\ line fell into our ALMA bandpass tuning. For other targets in our sample, our observations likely missed the galaxy's \cii\ line. Our $z\sim2$ galaxies follow the \cii-deficit relation observed for nearby (U)LIRGs, as found by several other $z\sim2-3$ studies.  
    \item As found in previous studies, our $z\sim2$ galaxies and other high-$z$ samples show decreasing \lsix/\lir\ with \lir. Star-forming galaxies at $z\sim2$ have more PAH emission per unit \lir\ compared to low$-z$ star-forming galaxies of comparable \lir; however, this offset disappears when comparing \lsix/\lir\ in all galaxies as a function of IR surface density. 
    \item We explore the balance of heating and cooling in the ISM by looking at the ratio of \cii\ to PAH luminosity. For nearby (U)LIRGs, this ratio is relatively tight as a function of \lir. Our $z\sim2$ galaxies are low relative to this relation. This may be because of warmer environments, suppressed photoelectric efficiencies in PDR gas, and/or the importance of cooling from other far-IR lines such as \oi\ at $z\sim2$. GS IRS61, the galaxy with the lowest \cii/PAH, shows evidence for high PAH ionization, consistent with inefficient gas heating in PDR regions. 
\end{enumerate}

We caution that our study shows that \cii\ and PAH emission may not have a simple relation to \lir, and therefore SFR, in $z\sim2$ dusty star-forming galaxies. Further observations are needed to validate our results and test the ideas of warmer dust environments and additional cooling channels. These can be obtained by getting more \cii\ detections of galaxies with existing PAH measurements from \textit{Spitzer/IRS} or from future programs tracing the mid-IR and far-IR lines with \textit{JWST} and ALMA. 
\newline

\acknowledgments
{\small We are very grateful to the referee for the detailed comments and suggestions that significantly improved this work. J.M. and A.P. acknowledge D. Brisbin and C. Ferkinhoff for assistance in interpreting ZEUS \cii\ results. J.M. acknowledges I. Yoon, J. Braatz, and J. Thorley for their insights into ALMA data reduction. We thank J.D. Smith for his helpful discussion. T.D-S. acknowledges support from the CASSACA and CONICYT fund CAS-CONICYT Call 2018. Support for this work was provided by the NSF through award SOSPA5-008 from the NRAO, and through the Massachusetts Space Grant Consortium (MASGS). The National Radio Astronomy Observatory is a facility of the National Science Foundation operated under cooperative agreement by Associated Universities, Inc. This paper makes use of the following ALMA data: ADS/JAO.ALMA\#2017.1.01347.S. ALMA is a partnership of ESO (representing its member states), NSF (USA) and NINS (Japan), together with NRC (Canada), MOST and ASIAA (Taiwan), and KASI (Republic of Korea), in cooperation with the Republic of Chile. The Joint ALMA Observatory is operated by ESO, AUI/NRAO and NAOJ. 
}


\newpage 
\appendix
\section{PAH-Derived redshifts and luminosities \label{pahmodel}} 

In this section, we describe our method for measuring the PAH redshifts and luminosities that employs MCMC to fully capture the uncertainties. \textit{Spitzer IRS} mid-IR spectra are shown in Figure \ref{irsSpectra}, which we use to calculate the redshift probability distribution function $p(z)$ and PAH line luminosities for galaxies in our sample. Rest frame mid-IR wavelengths are host to a diverse range of spectral features from rotational lines of molecular hydrogen to bending and stretching modes of PAH molecules. In the low SNR regime characteristic of high-redshift observations, only the brightest PAH features remain distinctly observable. These features are intrinsically broad with intensities $I_\nu^{(r)}$ well-fit by Lorentzian (Drude) profiles: 
\begin{equation}
    I_\nu^{(r)} = \frac{b_r\gamma_r^2}{(\lambda/\lambda_r - \lambda_r/\lambda)^2 + \gamma_r^2 )}
    \label{Drude}
\end{equation}
where following the convention of \cite{Smith2007}, $r$ specifies a given PAH complex with central wavelength $\lambda_r$, fractional full width at half maximum (FWHM) $\gamma_r$ and central intensity $b_r$. Lorentzian profiles are the theoretical spectrum for a classical damped harmonic oscillator, and carry more power in their extended wings compared to a Gaussian. As a result, individual line emission is difficult to separate from adjacent PAH features, as well as any underlying stellar and dust continuum (see \citealt{Smith2007} for examples at low-redshift). 

Owing to the number of blended line profiles between $5-15\mu$m, PAH flux densities in this wavelength domain are sensitive to the measurement technique (see \citealt{Smith2007} for a thorough analysis). In particular, how the continuum around each PAH feature is estimated can lead to variations in measured line fluxes and equivalent widths by up to a factor of four \citep{Sajina2007,Smith2007, Pope2008}. For this reason, we focus our analysis on the 6.2$\,\mu$m and 11.3$\,\mu$m PAH luminosities, as these features are comparatively isolated from adjacent lines and trace the total PAH luminosity (\lpah) with low scatter in local and high-$z$ star-forming galaxies \citep{Smith2007,Pope2008}. 

\begin{figure*}[t!]
    \gridline{\fig{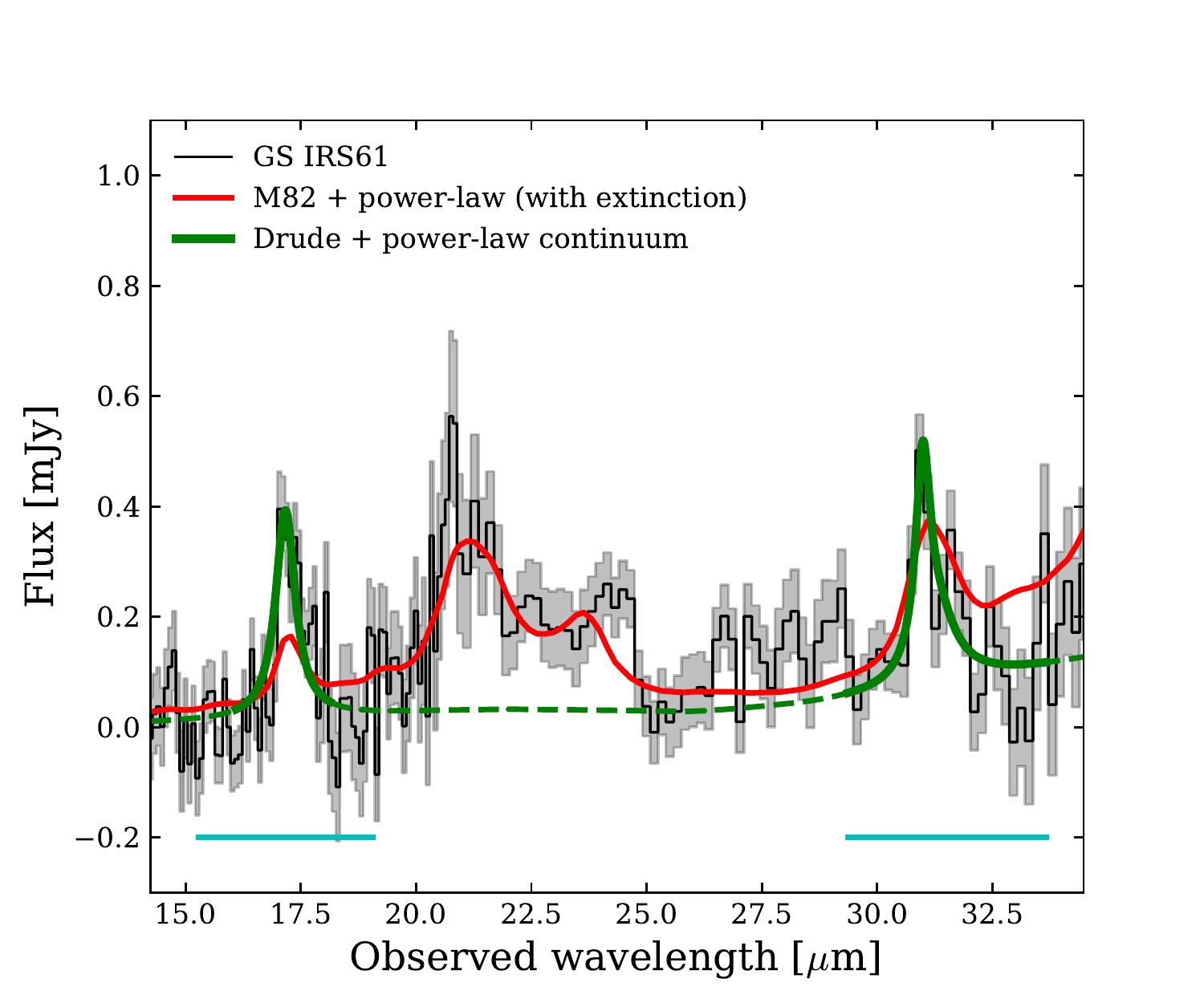}{0.45\textwidth}{}\hspace{-2em}
    \fig{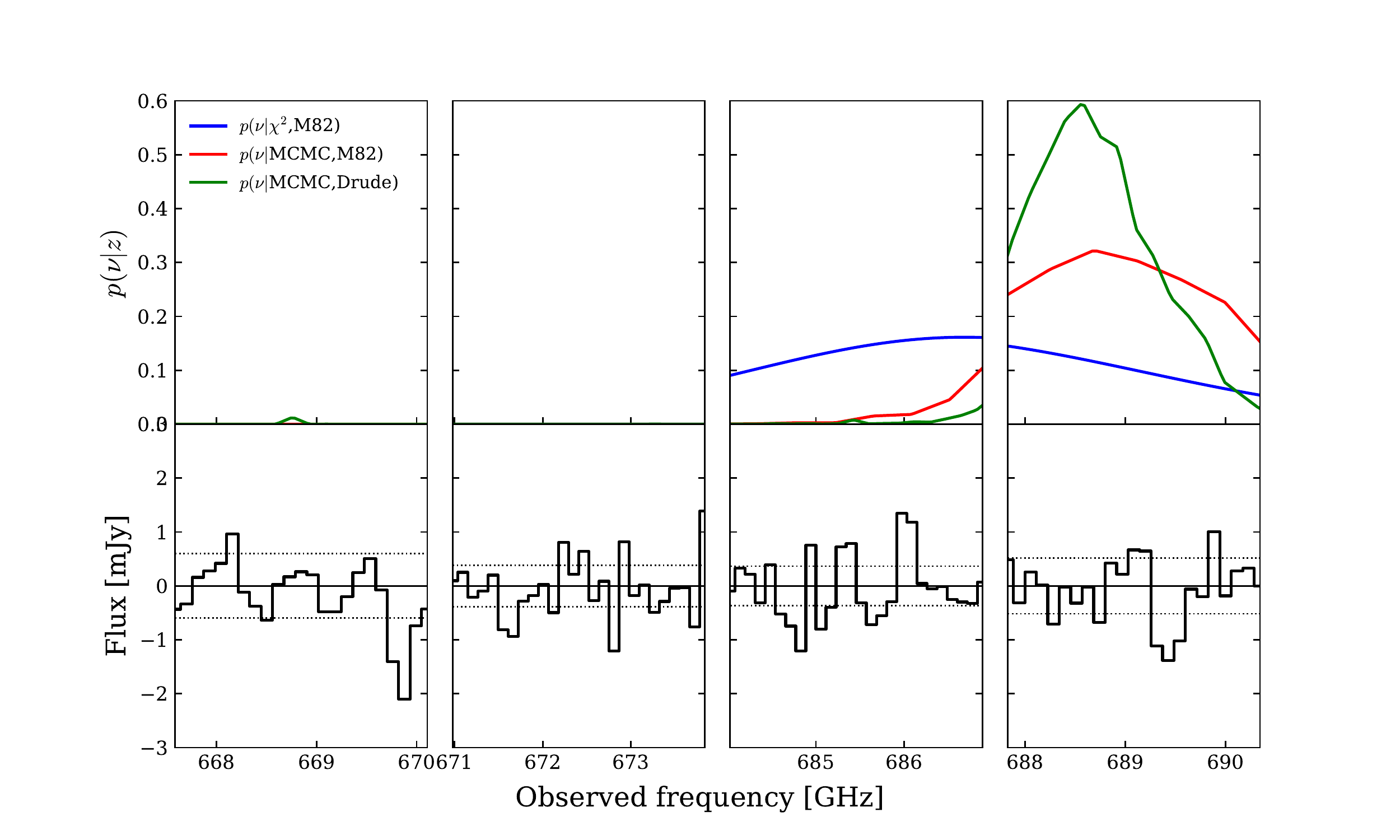}{0.63\textwidth}{}}
    \vspace{-2.5em}
    \caption{\small The importance of model selection in estimating redshifts from PAH spectra. \textit{(Left)}: GS IRS61's \textit{Spitzer IRS} spectrum. Over-plotted in green is our best fit Lorentzian model to the $6.2\mu$m and $11.3\mu$m PAH complexes. Shown in red is the AGN-SFG decomposition model of \cite{Kirkpatrick2015}, which we re-fit to GS IRS61's spectrum using the same MCMC package as was used in fitting the other model. Regions included in the fit are shown with a solid line, whereas dashed lines indicate wavelengths masked from each model. \textit{(Right)}: Bottom panels show GS IRS61's ALMA Band 9 spectrum, separated into regions of contiguous baseband coverage. Dotted black lines correspond to $\pm1\sigma$. Upper panels show $p(\nu|z)$, the probability of \cii\ being redshift to each observed frequency, for different fitting methods. Red and green lines correspond to the MCMC-derived redshift posteriors from fits in the \textit{Left} panel. The blue line shows a Gaussian approximation of $p(\nu|z)$ from $\chi^2$ minimization fits with the AGN-SFG decomposition model, originally used to plan the observations. From the green curve, we calculate the probability of having the \cii\ line fall in the ALMA bandpass to be 0.93 in GS IRS61.  \label{pzdemo}}
\end{figure*}

Inferring a redshift from PAH features at low SNR and low spectral resolution (R$\,\sim100$) is complicated by the many broad and blended PAH lines. Prior to our ALMA observations, redshifts were determined via the spectral decomposition model of \cite{Kirkpatrick2015} which fits mid-IR spectra with an AGN power-law component, a fixed star-forming galaxy PAH template, and dust extinction. This model works well for separating AGN and star formation components (i.e., calculating \fagn, Table \ref{sampsum}), but does not always reproduce observed PAH intensities as demonstrated in Figure \ref{pzdemo} (\textit{Left}). Peak emission at line center places the most constraint on a galaxy's systemic redshift. Therefore, we adopt a simpler model of Lorentzian profiles plus a power-law continuum to fit only the 6.2$\mu$m and 11.3$\mu$m PAH complexes. Using this technique, we leverage the relatively isolated lines to measure the target's redshift. In the rest-frame, our model is
\begin{equation}
    I_\nu = N_{pl} \lambda^{\alpha}e^{-\tau_{\nu,pl}} + \sum_{r} I_\nu^{(r)}(b_r|\lambda_r,\gamma_r)e^{-\tau_{\nu,pl}} 
    \label{model}
\end{equation}
where $N_{pl}$ is the power-law scale factor and $\alpha$ is the mid-IR spectral index. We assume a wavelength-dependent Milky Way dust attenuation law for the optical depth parameter $\tau_{\nu,pl}$ \citep{WeingartnerDraine2001}. This assumption has minimal to no impact on our results given that the primary purpose of the power-law component is to approximate continuum emission in the vicinity of each PAH feature. The second term in Equation \ref{model} sums over the various PAH complexes included in the fit, each described by a Lorentzian profile (Eq. \ref{Drude}). 


To fit for $b_r$ and $z$, we fix the central wavelengths of the $6.2\mu$m and $11.3\mu$m PAH features ($\lambda_r$) and $\gamma_r$ to their values derived in \cite{Smith2007}. This implicitly assumes comparable dust grain properties between high-$z$ (U)LIRGs and the inner kpc regions of galaxies from the \textit{Spitzer} Infrared Nearby Galaxies Survey (SINGS; \citealt{Kennicutt2003}) used to calibrate the PAH free-parameters. Many such features observed in low-$z$ star-forming galaxies are also seen in high-$z$ dusty systems, suggesting that the grain properties responsible for the intrinsically brightest PAH complexes (e.g., $6.2\mu$m, $7.7\mu$m, $8.6\mu$m, $11.3\mu$m) do not change between $z\sim2$ and today \citep{Pope2013, Kirkpatrick2015}. Although the 6.2 $\mu$m peak can shift by $\Delta\lambda_r\sim0.1$ $\mu$m from target to target in the Milky Way, these variations are related to the illumination source and relatively stable for both individual and averaged \ion{H}{2} regions and PDRs in the Milky Way which are expected to dominate mid-IR emission in star forming galaxies (Section 2.2.2 of \citealt{Tielens2008}, \citealt{Diedenhoven2004}). 

We fit our model (Eq. \ref{model}) to each \textit{IRS} spectrum using the Markov Chain Monte Carlo (MCMC) code \texttt{emcee}, an open-sourced package designed to minimize the number of tunable parameters embedded in a Markov Chain algorithm \citep{emcee}. We assume uniform priors on $N_{pl}$, $\alpha$, $\tau_{\nu,pl}$, $z$ and $b_r$, and restrict the fit to spectral domains around the $6.2\mu$m and $11.3\mu$m PAH features (see cyan horizontal lines in Figure \ref{irsSpectra}). The $6.2\mu$m and $11.3\mu$m features are unambiguous and readily identified by the code without confusion. 

Once the fits have been run, we marginalize over all free parameters and extract from each MCMC chain a redshift posterior probability distribution function $p(z)$. We quote the redshift that maximizes the likelihood function as $z_{IRS}$, and adopt uncertainties from the minimum and maximum redshifts within the 68th percentile of $p(z)$. Next, we use the local continuum around the 6.2 $\mu$m and 11.3 $\mu$m line to estimate \lsix\ and \lii, following measurement methods used in the literature for direct comparison with published values (e.g., \citealt{Uchida2000,Peeters2002,Pope2008,Pope2013}). Error bars on \lsix\ and \lii\ are derived using Monte Carlo analysis, whereby the observed spectrum is perturbed by pixel noise prior to re-calculating the line-flux and PAH feature luminosity. This process is repeated 1000 times, after which we quote the standard deviation of all iterations as the $1\sigma$ error. Final measurements and errors of $z_{IRS}$, \lsix, and \lii\ are provided in Table \ref{sampsum}. We note that silicate absorption at $9.7\,\mu$m can potentially impact the 11.3$\,\mu$m PAH feature shape and luminosity. There is little evidence for strong silicate absorption in the spectral decomposition shown in Fig. \ref{irsSpectra}; however, the low SNR data is consistent with optical depths of the 9.7$\,\mu$m feature $\tau_{9.7}\approx0-2$. At this opacity, the 11.3$\,\mu$m PAH feature strength is decreased by a factor of 1.4 at most \citep{Smith2007}, which is within the uncertainty of our measurements of \lii.


\section{Calculating the probability of observing \cii\ from total bandpass coverage \label{appendix:pz}}

In this section, we consider the uncertainties on the redshifts coupled with the ALMA bandpass to calculate the probabilities of observing \cii\ for each galaxy in our sample. This analysis is crucial before one can measure upper limits on \cii\ from ALMA data containing frequency gaps in baseband coverage. In designing the ALMA Cycle 5 observations, redshifts were determined for each source by fitting a single star-forming PAH template to each galaxy. As demonstrated by Figure \ref{pzdemo} (\textit{Left}), this method insufficiently matches the brightest PAH emission compared to a Lorentzian profile technique. While both fitting approaches estimate a redshift within $\pm1\sigma$ of each other corresponding to $ \Delta z = 0.014$ on average, differences on the order of $\Delta z\sim0.01$ can shift the \cii\ line in or out of the ALMA spectral windows at the highest frequencies. For this reason, our observations may have missed \cii\ in some galaxies. 

To quantify $p(l|\mathrm{ALMA},\Delta z)$, we take redshift posterior probability distributions from our MCMC fits to the IRS spectra (Section 3.1, Fig. \ref{irsSpectra}) and from these, compute $p(\nu_{\mathrm{\tiny [C II]}}|z)$: the probability \cii\ would be redshifted to a given frequency. Next, we integrate $p(\nu_{\mathrm{\tiny [C II]}}|z)$ first over all frequencies, and then over the frequency domain covered by our bandpass tunings which is typically $\sim10$ GHz between $632-687$ GHz not counting gaps between individual spectral windows. Thus, we quantitatively derive  $p(l|\mathrm{ALMA},\Delta z)$ according to the following prescription: 
\begin{equation}
     p(l|\mathrm{ALMA},\Delta z)\equiv \frac{\sum_i \int_{\mathrm{min}(\nu_i)}^{\mathrm{max}(\nu_i)} p(\nu_{\mathrm{\tiny [C II]}}|z) d\nu}{\int_{-\infty}^{+\infty} p(\nu_{\mathrm{\tiny [C II]}}|z) d\nu}
    \label{pz}
\end{equation}
where the summation treats each ALMA spectral window independently and avoids gaps in wavelength coverage. Figure \ref{pzdemo} (\textit{Right}) graphically demonstrates this technique for GS IRS61, the only galaxy in our observations where $p(l|\mathrm{ALMA},\Delta z)>90\%$. Estimates of $p(l|\mathrm{ALMA},\Delta z)$ for all other targets are given in Table \ref{almacont}. 

Additional redshift constraint from rest-frame optical spectroscopy can be used to improve the estimate of $p(l|\mathrm{ALMA},\Delta z)$. In principle, we would multiply optical redshift posteriors with our MCMC-derived $p(z)$ and integrate the product. We checked for optical spectroscopic redshifts by matching to catalogs from 3D-\textit{HST} grism \citep{Momcheva2016}, VLT/FORS-2 \citep{Vanzella2008}, VANDELS \citep{vandels}, MUSE GTO surveys, and ALESS \citep{Danielson2017}. GS IRS50 and GS IRS58 have grism spectroscopic redshifts consistent with \zirs\ but with higher uncertainty. GS IRS20 and GS IRS61 have C-grade VLT/FORS-2 spectra, and grism redshifts completely inconsistent with the PAH features in both galaxies by $\Delta z = 0.2-0.3$, greater than 10 times the uncertainty on their PAH-derived redshifts. GS IRS46 and GS IRS52 do not have optical spectra. In summary, no significantly accurate optical spectroscopic redshifts ($\Delta z<0.01$) consistent with \zirs\ were found that changed our results using only PAH fits. 


\begin{longrotatetable}
\begin{deluxetable}{lcccccccccccc}
    \tablecaption{Existing \textit{Spitzer} and \textit{Herschel} Photometry\label{phot}}
    \tablecolumns{13}
    \tabletypesize{\scriptsize}
    \tablewidth{\paperheight}
    \tablehead{Target &   3.6$\mu$m & 4.5$\mu$m & 5.8$\mu$m & 8$\mu$m & 16$\mu$m & 24$\mu$m & 70$\mu$m & 100$\mu$m & 160$\mu$m & 250$\mu$m \tablenotemark{a}& 350$\mu$m\tablenotemark{a} & 500$\mu$m\tablenotemark{a} \\
                      & [$\mu$Jy]  & [$\mu$Jy] & [$\mu$Jy] & [$\mu$Jy] & [$\mu$Jy] & [$\mu$Jy] & [$\mu$Jy] & [mJy] & [mJy] & [mJy] & [mJy] & [mJy] }
    \startdata 
    GS IRS20 & $16.91\pm0.05$& $19.89\pm0.07$& $21.59\pm0.32$&$13.42\pm0.35$&\---&$275.0\pm6.0$&$3170.0\pm590.5$&$8.32\pm0.56$&$20.15\pm1.6$& $24.78\pm2.3$& $23.82\pm3.08$&$12.49\pm5.08$\\[1ex]
    GS IRS46 & $68.87\pm0.06$& $51.16\pm0.08$&$42.72\pm0.47$&$25.45\pm0.45$&\---&$378.0\pm6.6$&\---&\---&\---&$13.05\pm2.8$& $15.11\pm4.7$&$13.0\pm2.0$\\ [1ex]
    GS IRS50 & $14.9\pm0.03$& $16.7\pm0.04$& $15.88\pm0.22$&$10.49\pm0.23$&$55.6\pm6.6$&$227.0\pm8.1$&\---&$1.12\pm0.13$&$3.62\pm1.05$&$5.83\pm2.84$&$6.06\pm2.44$&$4.03\pm3.44$\\ [1ex]
    GS IRS52 & $9.93\pm0.05$& $13.47\pm0.06$&$13.15\pm0.33$&$10.7\pm0.33$&$60.9\pm8.1$&$240.0\pm8.6$&\---&\---&\---& $10.06\pm3.26$&$9.05\pm4.4$& $6.86\pm3.66$\\ [1ex]
    GS IRS58 & $18.45\pm0.04$& $21.48\pm0.06$&$18.92\pm0.27$&$12.82\pm0.31$&\---& $243.0\pm8.7$&\---&$1.27\pm0.14$&$3.51\pm0.99$&$8.85\pm2.28$ & $11.15\pm2.14$&$5.73\pm2.88$\\ [1ex]
    GS IRS61 & $16.21\pm0.04$& $18.73\pm0.05$&$17.05\pm0.25$&$13.37\pm0.27$&$59.6\pm5.0$& $245.0\pm8.7$&\---&$3.21\pm0.16$ &$5.56\pm1.08$&$5.76\pm2.2$& $3.44\pm2.58$ &$4.96\pm2.8$
    \enddata
    \tablenotetext{a}{Confusion noise for 250 $\mu$,, 350 $\mu$m and 500 $\mu$m is $\sim5$ mJy (see \citealt{Nguyen2010} for exact values) and has been included in all SED fits. }
\end{deluxetable}
\end{longrotatetable}


\bibliography{references.bib}

\end{document}